\documentclass{aa}  

\usepackage{graphicx}
\usepackage{txfonts}
\usepackage{color}

\begin{document}

   \title{HADES RV programme with HARPS-N at TNG\thanks{Based on observations made with the Italian {\it Telescopio Nazionale Galileo} (TNG) operated by the {\it Fundaci\'on Galileo Galilei} (FGG) of the {\it Istituto Nazionale di Astrofisica} (INAF) at the  {\it Observatorio del Roque de los Muchachos} (La Palma, Canary Islands, Spain).}
\fnmsep \thanks{
Table~\ref{M059_log} is only available in electronic form
at the CDS via anonymous ftp to cdsarc.u-strasbg.fr (130.79.128.5)
or via http://cdsweb.u-strasbg.fr/cgi-bin/qcat?J/A+A/   
   }
 }

   \subtitle{XIV. A candidate super-Earth orbiting the M-dwarf GJ 9689 with a period close to half the stellar rotation period}

   \author{J. Maldonado\inst{1}  
          \and A. Petralia\inst{1}
          \and M. Damasso   \inst{2}
          \and M. Pinamonti \inst{2}
          \and G. Scandariato \inst{3}
	  \and E. Gonz\'alez-\'Alvarez \inst{4}
	  \and L. Affer \inst{1}
          \and G. Micela \inst{1}
	  \and  A. F. Lanza  \inst{3}
	  \and  G. Leto \inst{3}
	  \and  E. Poretti   \inst{5,6}
	  \and  A. Sozzetti  \inst{2}
	  \and  M. Perger    \inst{7,8}
	  \and  P. Giacobbe  \inst{2}
	  \and  R. Zanmar S\'anchez \inst{3}
	  \and  A. Maggio \inst{1}
          \and J. I. Gonz\'alez Hern\'andez \inst{9,10}
	  \and R. Rebolo  \inst{9,10}
	  \and I. Ribas     \inst{7,8}
	  \and A. Su\'arez-Mascare\~{n}o \inst{9,10}
	  \and B. Toledo-Padr\'on \inst{9,10}
          \and  A. Bignamini \inst{11}
	  \and  E. Molinari  \inst{12}
	  \and E. Covino \inst{13}
          \and R. Claudi \inst{14}
	  \and S. Desidera \inst{14}
	  \and E. Herrero \inst{7,8}
	  \and J. C. Morales \inst{7,8}
	  \and I. Pagano \inst{3}
	  \and G. Piotto \inst{15}
          }

   \institute{INAF - Osservatorio Astronomico di Palermo,
              Piazza del Parlamento 1, 90134 Palermo, Italy\\
              \email{jesus.maldonado@inaf.it}
            \and INAF - Osservatorio Astrofisico di Torino, Via Osservatorio 20, 10025 Pino Torinese, Italy 
	    \and INAF - Osservatorio Astrofisico di Catania, Via S. Sofia 78, 95123, Catania, Italy
	    \and Centro de Astrobiolog\'ia (CSIC-INTA), Carretera de Ajalvir km 4, 28850 Torrej\'on de Ardoz, Madrid, Spain
	    \and 
	    Fundaci\'on Galileo Galilei-INAF, Rambla Jos\'e Ana Fern\'andez P\'erez 7, 38712 Bre\~{n}a Baja, TF, Spain
	    \and 
	    INAF - Osservatorio Astronomico di Brera, Via E. Bianchi 46, 23807 Merate, Italy
	    \and Institut de Ci\'encies de l'Espai (ICE, CSIC), Campus UAB, Carrer de Can Magrans s/n, 08193 Bellaterra, Spain
	    \and Institut d'Estudis Espacials de Catalunya (IEEC), 08034 Barcelona, Spain
            \and Instituto de Astrof\'isica de Canarias, 38205 La Laguna, Tenerife, Spain
            \and Universidad de La Laguna, Departamento Astrof\'isica, 38206 La Laguna, Tenerife, Spain
	    \and INAF - Osservatorio Astronomico di Trieste, Via Tiepolo 11, 34143 Trieste, Italy
	    \and INAF - Osservatorio Astronomico di Cagliari \& REM, Via della Scienza, 5, 09047 Selargius CA, Italy
	    \and INAF - Osservatorio Astronomico di Capodimonte, Salita Moiariello 16, 80131 Napoli, Italy
	    \and INAF - Osservatorio Astronomico di Padova, vicolo dell'Osservatorio 5, 35122 Padova, Italy
	    \and Dipartimento di Fisica e Astronomia Galileo Galilei, Vicolo Osservatorio 3, 35122 Padova, Italy
             }

   \date{Received ; accepted }

 
  \abstract
   {It is now well-established that small, rocky planets are common around low-mass stars.
    However, the detection of such planets is challenged by the short-term activity of the host stars.}
   {The HArps-N red Dwarf Exoplanet Survey (HADES) program is a long-term project at the Telescopio Nazionale Galileo aimed at the monitoring of nearby, early-type, M dwarfs, using the HARPS-N spectrograph to search for small, rocky planets.} 
   {A total of 174 HARPS-N spectroscopic observations of the M0.5V-type star GJ 9689 taken over the past seven years have been analysed.
   We combined these data with photometric measurements to disentangle signals related to the stellar activity of the star from possible
   Keplerian signals in the radial velocity data. We run an MCMC analysis, applying Gaussian Process regression techniques to model 
   the signals present in the data.}
   {We identify two periodic signals in the radial velocity time series, with periods of 18.27 d, and 39.31 d.
    The analysis of the activity indexes, photometric data, and wavelength dependency of the signals reveals that the
    39.31 d signal corresponds to the stellar rotation period. On the other hand, the 18.27 d signal shows no relation to
    any activity proxy  or the first harmonic of the rotation period.  We, therefore, identify it as a genuine Keplerian signal.
    The best-fit model describing   the newly
    found planet, GJ 9689 b, corresponds to an orbital period P$_{\rm b}$ = 18.27 $\pm$ 0.01 d,
    and a minimum mass M$_{\rm P}\sin i$ = 9.65 $\pm$ 1.41 M$_{\oplus}$. 
   }
   {}


   \maketitle
%
\section{Introduction}

Starting with the first exoplanet discoveries \citep{1992Natur.355..145W,1995Natur.378..355M}
astronomers have succeeded in unravelling an astonishing diversity of planetary systems,
compositions, and host stellar properties 
\citep[e.g.][]{2007ARA&A..45..397U,2010exop.book..191C,2013Sci...340..572H,2018exha.book.....P}.

Low-mass stars or M dwarfs constitute the most common stellar component in the Milky Way \citep[e.g.][]{2000ARA&A..38..337C}.
However, unlike their FGK counterparts, results regarding the properties of planets
around M dwarfs still focus on small sample sizes and suffer from the intrinsic difficulties
in the spectral characterisation of M dwarfs.
 In spite of their high levels of
stellar variability and their intrinsic faintness at optical wavelengths,
it is now widely
accepted that low-mass stars are optimal targets around which to search for small rocky planets
\citep[e.g.][]{2013ApJ...767...95D,2015ApJ...807...45D} 
and  a large number of both radial velocity 
as the HET and HPF M dwarf surveys \citep[][]{2003AJ....126.3099E,2018AAS...23124645W}, 
the HARPS-GTO programme \citep{2013A&A...549A.109B}, the HADES survey \citep{2016A&A...593A.117A},
CARMENES \citep[e.g.][]{2019A&A...627A..49Z}, SPIRou \citep[e.g.][]{2021MNRAS.502..188K},
or SOPHIE \citep[e.g.][]{2019A&A...625A..18H},
and photometric surveys 
like MEarth \citep[][]{2008PASP..120..317N}, APACHE \citep{2013EPJWC..4703006S}, 
TRAPPIST \citep{refId0}, or TESS \citep[e.g.][]{2021AJ....161..247F}
are currently focused on M dwarfs.

At the time of writing there are $\sim$ 130 known radial velocity planets around M dwarfs listed in The Extrasolar
Planets Encyclopaedia \citep{2011A&A...532A..79S}\footnote{http://exoplanet.eu/, as checked in December 2020},
and several interesting trends regarding the properties of planets around M dwarfs have already been discussed.
It has been established that, unlike their solar-type counterparts, the frequency of gas-giant planets orbiting
low-mass stars is low \citep{2003AJ....126.3099E,2006ApJ...649..436E,2006PASP..118.1685B,2007A&A...474..293B,2008PASP..120..531C,2010PASP..122..149J}.
On the other hand, as found for FGK stars, small planets orbiting
M dwarfs seem to be very abundant, most of them in multi planet
systems.
 The HARPS-M dwarf survey reports a 36\% occurrence for 
 super-Earths (M$_{\rm p} \sin i$ $<$ 10 M$_{\oplus}$) in short periods
 ($P$ < 10 d) and 52\% for 10 days $<$ $P$ $<$ 100 d \citep{2013A&A...549A.109B}.
 \cite{2014MNRAS.441.1545T} estimates a frequency of low-mass planets around M dwarfs of one planet per
 star, possibly even greater. 
 Results from the {\sc Kepler} mission
 suggest that the frequency of planets (with periods below 50 days) around stars seems to increase as 
 we move from F stars towards the M dwarf type \citep{2012ApJS..201...15H}. 
 Along this line, \cite{2012ApJ...753...90M}
 report an occurrence rate of super-Earths 
 with periods lower than 50 days of 36 $\pm$ 8\% around late-K to early M dwarfs. 

 The frequency of planets around M dwarfs seems to follow the same stellar mass and metallicity
trends than FGK stars. That is, the frequency of gas-giant planets is a function of the stellar metallicity as well as of the stellar mass.
On the other hand, the frequency of low-mass planets does not depend on the metallicity content or
the mass of the stellar host \citep{2007A&A...474..293B,2009ApJ...699..933J,2010A&A...519A.105S,2012ApJ...748...93R,2012ApJ...747L..38T,2013A&A...551A..36N,2014ApJ...781...28M,2016MNRAS.461.1841C,2020A&A...644A..68M}.
While most studies have focused only on the iron content or metallicity,
in a recent work, \cite{2020A&A...644A..68M} show, for the first time, that there are no differences in the abundance distribution of elements other than iron
between M dwarfs with and without known planets.

 The detection of truly Earth twins via the Doppler technique requires a precision
 in the radial velocity measurements of the order of 0.1 ms$^{\rm -1}$. While
 a new generation of ultra high-resolution spectrographs is coming to the foe 
 \citep[e.g. ESPRESSO][]{2021A&A...645A..96P},
 it is now clear that the major challenge to high Doppler precision is
 not the instrumental precision but the star itself. 
 Stellar activity might produce line profile variations that skew the peak of 
 a spectral line, leading to a velocity change in the star that 
 can be (mis)-interpreted as Keplerian in nature. 
 \citep{1997ApJ...485..319S,2000A&A...361..265S,2004AJ....127.3579P,2005PASP..117..657W,2007A&A...473..983D,2011A&A...527A..82D}.
 Some techniques such as an optimal schedule of the observations or the
 use of spectroscopic indicators of activity  might be used to mitigate the effects of stellar oscillations,
 granulation or even the long-term activity \citep{2011A&A...527A..82D,2011A&A...525A.140D}. 
 However, the short-term stellar activity (due to evolution and decay of active regions)
 has a characteristic timescale comparable with the stellar rotation period \citep[e.g.][]{2017A&A...598A..28S}. 
 Disentangling ``true'' keplerian signals from stellar variations is highly complex
 and requires the use of complementary and imaginative approaches like
 red-noise models, 
 or Gaussian process regression \citep[see e.g.][and references therein]{2017A&A...598A.133D}.

 Within the framework of the HArps-n red Dwarf Exoplanet Survey (HADES) observing program
 \citep{2016A&A...593A.117A} we have started the radial velocity monitoring of a large sample of
 low-mass stars (spectral types K7-M3). HADES main goal is to explore the frequency and formation
 conditions of small, potentially habitable planets around early-M dwarfs.
 The development of techniques to ensure the optimal outcome
 of the survey is an additional goal of HADES, and it includes target characterisation \citep{2015A&A...577A.132M,2020A&A...644A..68M},
 optimal scheduling \citep{2017A&A...598A..26P},
 or detailed activity studies \citep{2017A&A...598A..27M,2017A&A...598A..28S,2018A&A...612A..89S,2019A&A...624A..27G}. 
 HADES has already succeeded in discovering several super-Earth exoplanets,
 with masses ranging from 2.5 M$_{\oplus}$ to 9 M$_{\oplus}$
 \citep{2016A&A...593A.117A,2017arXiv170506537S,Manuel,2018A&A...617A.104P,2019A&A...625A.126P,2019A&A...622A.193A,2019A&A...624A.123P,
 2021A&A...648A..20T,2021arXiv210309643G}.

  In this paper we present the discovery of a candidate super-Earth 
  (M$_{\rm p}\sin i$ $\sim$ 9.65 M$_{\oplus}$) planet
  orbiting around the early-M dwarf GJ 9689 with a period of 18.27 d.
  The detection of this planet has been challenging, as the proposed planetary period is
  very close to half
  the stellar rotation period (39.31 d).
  This fact makes the GJ 9689 b planet an interesting case of study. 
  This paper is organised as follows. 
  Section~\ref{host_star} reviews the stellar properties of GJ 9689.
  The spectroscopic data is presented in Sect.~\ref{spect_obs}
  while Sect.~\ref{time_series} describes the analysis of the radial velocity data.
  The origin of the radial velocity variations found in GJ 9689 is discussed at length
  in Sect.~\ref{origin} where activity indicators, photometry, and CCF diagnostics
  are analysed. Section~\ref{modelling} describes the modelling of the radial 
  velocity variations. The results are set in the context of planetary systems
  in Sect.~\ref{others}.
  Our conclusions follow in Sect.~\ref{conclusions}.

\section{The host star}\label{host_star}

 GJ 9689 is an M0.5 dwarf located at a distance of 30.69 $\pm$ 0.01 pc \citep{2020yCat.1350....0G}
 from the Sun. Its main stellar properties are listed in Table~\ref{physical_properties}.
 Basic stellar parameters (effective temperature, spectral type,
 mass, radius, surface gravity, and luminosity)
 are from \cite{2020A&A...644A..68M}. They are 
 computed following the procedures described in
 \cite{2015A&A...577A.132M}\footnote{https://github.com/jesusmaldonadoprado/mdslines}
 which make use of the same spectra used in this work for the radial velocity analysis.
 In brief, effective temperatures and spectral types are determined from
 ratios of pseudo-equivalent widths of spectral features calibrated using
 stars with interferometric estimates of their radii and spectral-type standards.
 By studying a large sample of early-M dwarfs, \citet{2015A&A...577A.132M} also 
 provide empirical calibrations for the stellar mass, radius, and surface gravity
 as a function of the stellar metallicity and effective temperature.
 Stellar metallicity is computed by a methodology  based on the use of a principal component analysis and sparse Bayesian methods
 which also allows the determination of abundances of other elements different from iron\footnote{https://github.com/jesusmaldonadoprado/mdwarfs\_abundances} 
 \citep{2020A&A...644A..68M}.
 These parameters are listed for the whole HADES sample in \cite{2017A&A...598A..27M}
 and for a large sample of M dwarfs in current radial velocity searches in
 \cite{2020A&A...644A..68M}.

 Galactic spatial-velocity components $(U, V, W)$
 are 
 computed
 from the radial velocities, together with {\it Gaia} parallaxes
 and proper motions \citep{2020yCat.1350....0G}  
 following the procedure described in \cite{2001MNRAS.328...45M} and 
 \cite{2010A&A...521A..12M}.
 GJ 9689 is classified as transition (thin/thick disk) star
 applying the methodology described in
 \cite{2003A&A...410..527B,2005A&A...433..185B}.
 No comoving objects seems to be present after carefully checking the available data in the
 {\it Gaia} EDR3 catalogue.  

\begin{table}[!htb]
\centering
\caption{Physical properties of GJ 9689}
\label{physical_properties}
\begin{tabular}{lrl}
\hline\noalign{\smallskip}
 Parameter                 &  Value               &  Notes \\
\hline
 $\alpha$ (ICRS J2000)     & 20:13:51.8	          &        \\
 $\delta$ (ICRS J2000)     & +13:23:20            &        \\ 
\hline
 T$_{\rm eff}$ (K)         &  3836 $\pm$ 69       &  a  \\ 
 Spectral Type             &  M0.5                &  a  \\
 $[Fe/H]$ (dex)            &  0.05 $\pm$ 0.04    &  a  \\
 M$_{\star}$ (M$_{\odot}$) &   0.59 $\pm$ 0.06    &  a  \\
 R$_{\star}$ (R$_{\odot}$) &   0.57 $\pm$ 0.06    &  a  \\
 $\log g$ (cm s$^{\rm -2}$) &   4.69 $\pm$ 0.05    &  a  \\
 $\log (L_{\star}/L_{\odot})$ & -1.20 $\pm$ 0.09  &  a  \\
\hline
 v$\sin i$ (km s$^{\rm -1}$) & $<$ 1.47           & b  \\
 Age$^{\dag}$ (Gyr)                   & 8.9 $\pm$ 3.9      & a  \\
\hline
 $\pi$ (mas)                     &  32.5879  $\pm$ 0.0140    & c   \\
 $\mu_{\alpha}$ (mas/yr)         &  421.921  $\pm$ 0.014     & c   \\
 $\mu_{\delta}$ (mas/yr)         &   19.129  $\pm$ 0.015     & c   \\ 
 v$_{\rm rad}$ (km s$^{\rm -1}$) &   -67.701  $\pm$ 0.03     &    \\
 $U$ (km s$^{\rm -1}$)           &   -72.54  $\pm$ 0.02      &    \\
 $V$ (km s$^{\rm -1}$)           &   -42.24  $\pm$ 0.01      &    \\ 
 $W$ (km s$^{\rm -1}$)           &   -36.21  $\pm$ 0.02      &    \\
\hline  
 $V$ (mag)                 & 11.30             & d \\
 $(B-V)$ (mag)             & 1.365 $\pm$ 0.133 & d \\ 
 $(V-I)$ (mag)             & 1.60  $\pm$ 0.25  & d \\
\hline
 2MASS J (mag)             & 8.309 $\pm$ 0.029 & e \\ 
 2MASS H (mag)             & 7.633 $\pm$ 0.021 & e \\
 2MASS K$_{\rm S}$ (mag)   & 7.468 $\pm$ 0.021 & e \\
\hline
\end{tabular}
\tablefoot{(a) \cite{2020A&A...644A..68M}; (b) \cite{2017A&A...598A..27M}; 
 (c) \cite{2020yCat.1350....0G}; 
 (d) \cite{1997ESASP1200.....E}; (e) \cite{2003yCat.2246....0C}.\\
 $^{\dag}$ By interpolation of parallaxes and stellar parameters within a grid of Yonsei-Yale isochrones, see (a) for details.}
\end{table}

\section{Spectroscopic observations}\label{spect_obs} 
%
 GJ 9689 has been observed during a period of almost seven and a half years,
 from BJD = 2456438  (May 26, 2013) to  BJD = 2459130 (October 7, 2020).
 A total of 174 HARPS-N observations
 were collected during this period. HARPS-N spectra cover the
 wavelength range 383-693 nm with a resolving power of $R$ $\sim$ 115000.
 Data were reduced using the latest version of the Data Reduction Software
 \citep[DRS V3.7,][]{2007A&A...468.1115L} which
 implements the typical corrections 
 involved in \'echelle spectra reduction, i.e. bias level, flat-fielding, 
 order extraction, wavelength calibration, and merge of individual orders.
 RVs are computed by 
 cross-correlating the spectra of the target star with an optimised
 binary mask \citep{1996A&AS..119..373B,2002A&A...388..632P}.
 For GJ 9689 the M2 mask was used.
 This procedure is known to have several shortcomings starting from
 the fact that a symmetric analytical function is used to fit the
 asymmetric CCF. Furthermore, the spectra of an M dwarf suffer from heavy blends
 resulting in side-lobes in the CCF that might affect the RV precision
 as well as the asymmetry indexes of the CCF \citep[e.g.][]{2020ExA....49...73R}. 
 In order to overcome these difficulties, RVs were also computed with the
 Java-based Template-Enhanced Radial velocity Re-analysis Application
 \citep[TERRA,][]{2012ApJS..200...15A}.
 TERRA measures the RVs by a least-square match of each observed spectrum to a 
 co-added high signal-to-noise (S/N) template
 spectra derived from the same observations. 
 We excluded from the analysis the bluest part of the spectra and
 only
 orders redder than the 22nd ($\lambda$ $>$  453 nm) were considered
  (note that HARPS-N spectra have a total of 66 \'echelle orders).

\section{Radial velocity time series analysis}\label{time_series}
\label{rv_time_series}
%

\begin{figure}[!htb]
\centering
\includegraphics[scale=0.60]{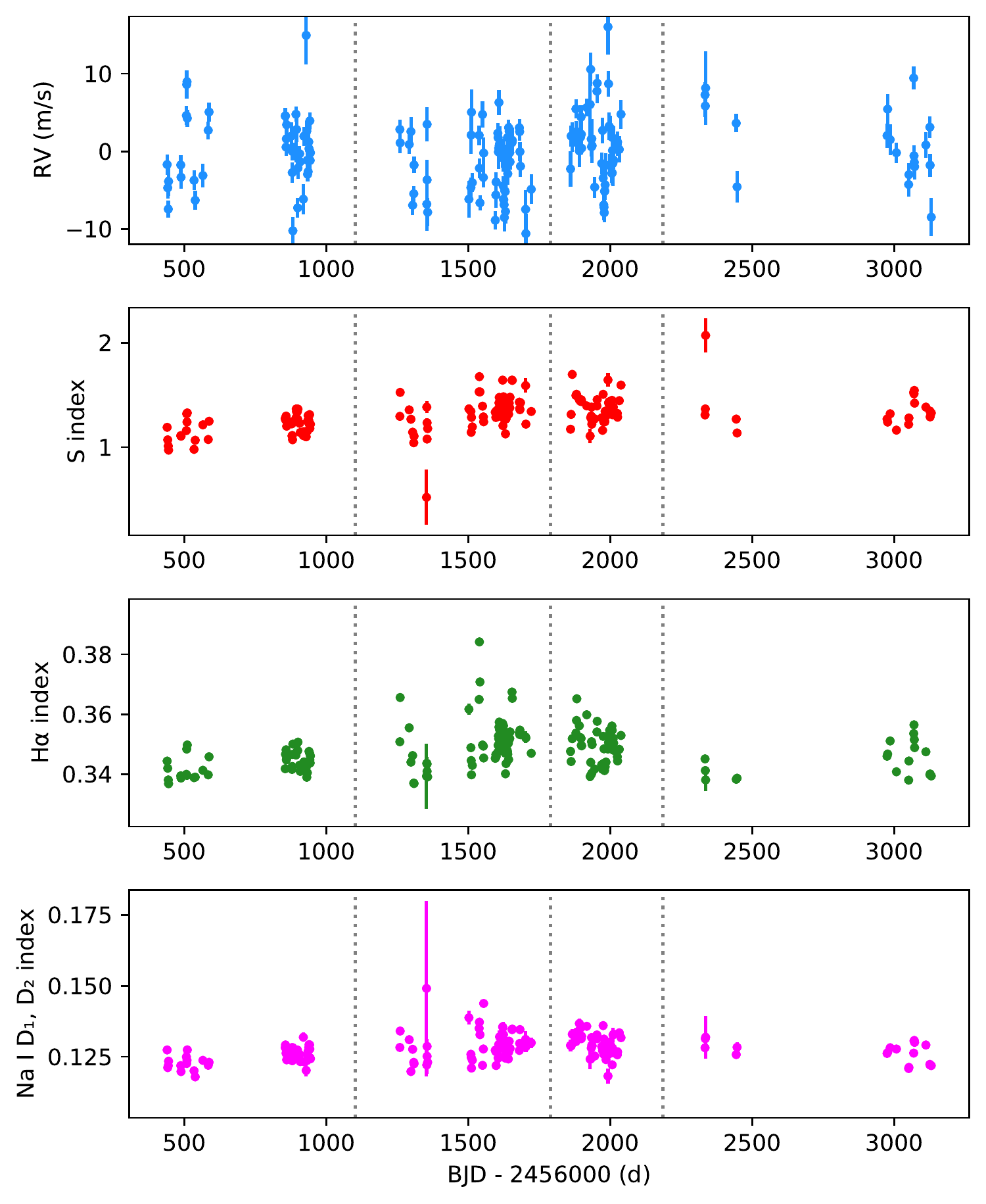}
\caption{
From top to bottom: Original radial velocity (derived with the TERRA pipeline), S-index, H$\alpha$-index, and Na~{\sc i} D$_{\rm 1}$, D$_{\rm 2}$-index time series for GJ 9689.
Vertical dotted lines indicate the  first three observing seasons as used in our study. 
}
\label{rv_series}
\end{figure}

 Figure~\ref{rv_series} (top) shows the RV (derived with the TERRA pipeline) time series of GJ 9689.
 The RV data show a rms of 4.59 m s$^{\rm -1}$, approx. 3 times
 the mean error of the measured RVs (1.55 m s$^{\rm -1}$)
 when the TERRA pipeline is used.
 If the RV data are 
 derived by the DRS then we obtain a rms of 5.35 m s$^{\rm -1}$ while the 
 mean error of the measured RVs is 2.94 m s$^{\rm -1}$.

 In a recent work, \citet[][]{2017A&A...598A..26P} perform
 a detailed comparison on the accuracy of the TERRA and DRS
 pipelines using the HADES spectra collected so far.
 Under the assumption that smaller
 rms of the RV measurements correspond
 to smaller RV noise rms, the authors conclude that
 TERRA RVs should be preferred.
 In particular, for GJ 9689, the rms of the RV measurements obtained with TERRA
 is around 1 ms$^{\rm -1}$ lower than the value derived by the DRS RVs. 
 The analysis presented in the following  
 refers to TERRA RVs. 
 
%
%
%
  
\begin{figure}[!htb]
\centering
\includegraphics[scale=0.65]{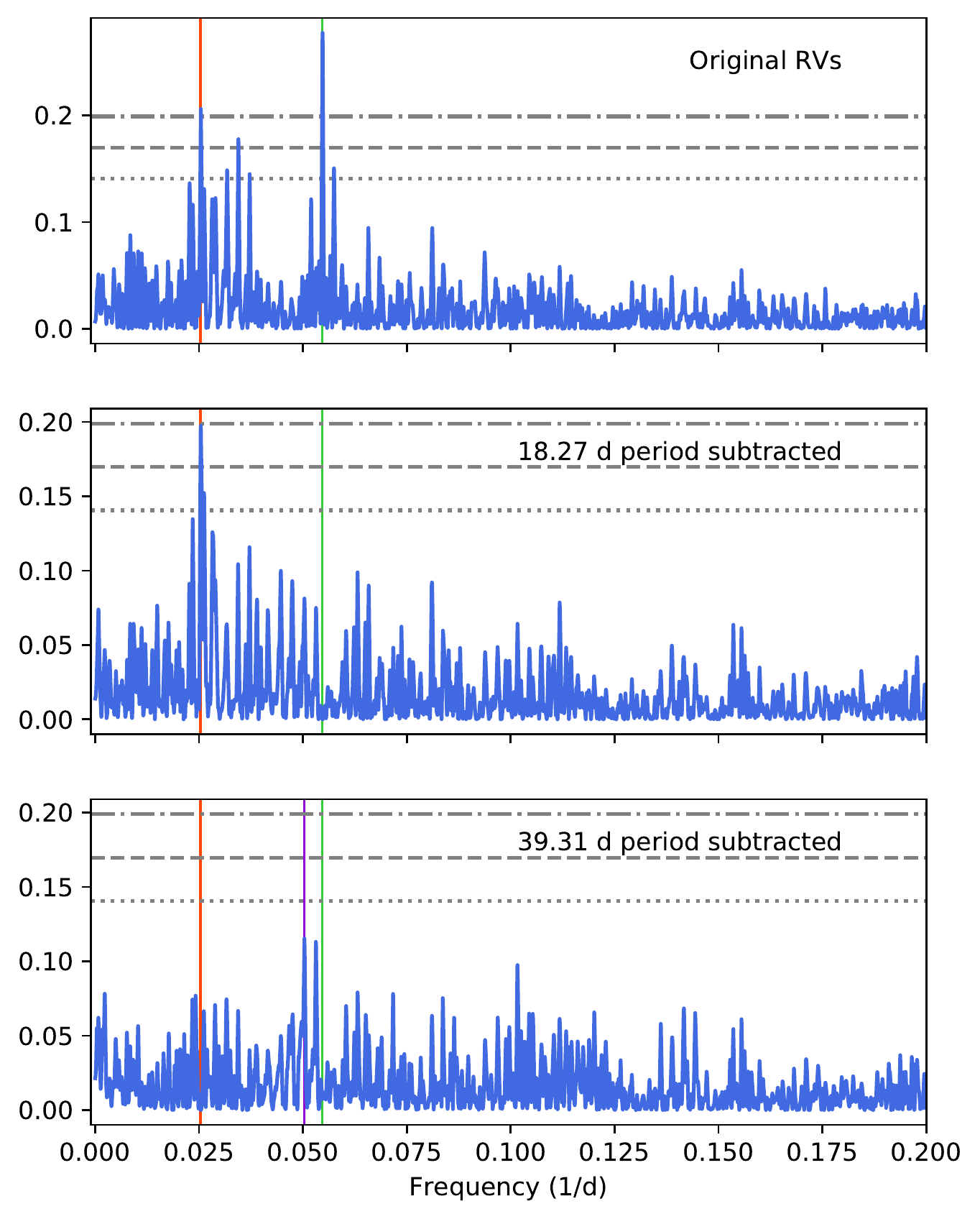}
\caption{Top: GLS periodogram of the TERRA RV measurements.
Middle: GLS periodogram after subtracting the 18.27 d period.
Bottom: GLS periodogram after subtracting the 18.27 and the 39.31 d signals. Values
corresponding to a FAP of 10\%, 1\%, and 0.1\% are shown with 
horizontal grey lines. The vertical red line indicates the period at 39.31 d
while the vertical green line shows the 18.27 d period.
The first harmonic of the 39.31 d signal is shown in violet. 
} 
\label{rv_series_periodogram}
\end{figure}

  A search for periodicities in the RV data was performed by using
  the generalised Lomb-Scargle periodogram \citep[GLS,][]{2009A&A...496..577Z}.
  The periodogram, see Figure~\ref{rv_series_periodogram},  
  identifies two significant frequencies. The
  peaks are found at 0.054734 $\pm$ 0.000039 d$^{\rm -1}$
  (period of 18.27 $\pm$ 0.01 d), 
  and 0.025436 $\pm$ 0.000059 d$^{\rm -1}$ which 
  corresponds to a period of 39.31 $\pm$ 0.10 d.  
  In order to test the significance of these frequencies a bootstraping analysis was done.
  A series of 10$^{\rm 4}$ simulations
  was performed. In each simulation the dates of the observed RVs were kept,
  but random RVs were constructed from the original ones, by drawing random values
  from normal distributions with means equals to the RV value and $\sigma$ equal to the RV error.
  Then, the simulated RVs were randomly
  shuffled.   
  The false alarm probability (FAP) of a given period is obtained as
  the fraction of simulated periodograms in which a peak with a
  periodogram power larger than the original period's power is
  found \citep[e.g.][]{2001A&A...374..675E}. Both signals are found to have
  a FAP below the 0.1\% threshold. 

%
  We performed a search for additional periods by subtracting
  in a sequential way the most prominent period until no significant
  signal is left, a procedure usually known as prewhitening.
  First of all, a sinusoidal function with period 18.27 d was fitted and subtracted.
  It can be seen from the periodogram, Figure~\ref{rv_series_periodogram}
  (middle panel), that the period at 39.31 d is still significant after the removal
  of the 18.27 d signal. 
  Once the 39.31 d signal is also subtracted, no significant periods
  remain in the periodogram analysis, Figure~\ref{rv_series_periodogram} 
  (bottom panel). We note that the peak with the highest power seems to be 
  the harmonic of the 39.31 d signal, as it is located
  at a frequency of 0.050364 $\pm$ 0.000075 d$^{\rm -1}$, which
  corresponds to a period of 19.86 $\pm$ 0.03 d.

\begin{figure}[!htb]
\centering
\includegraphics[scale=0.65]{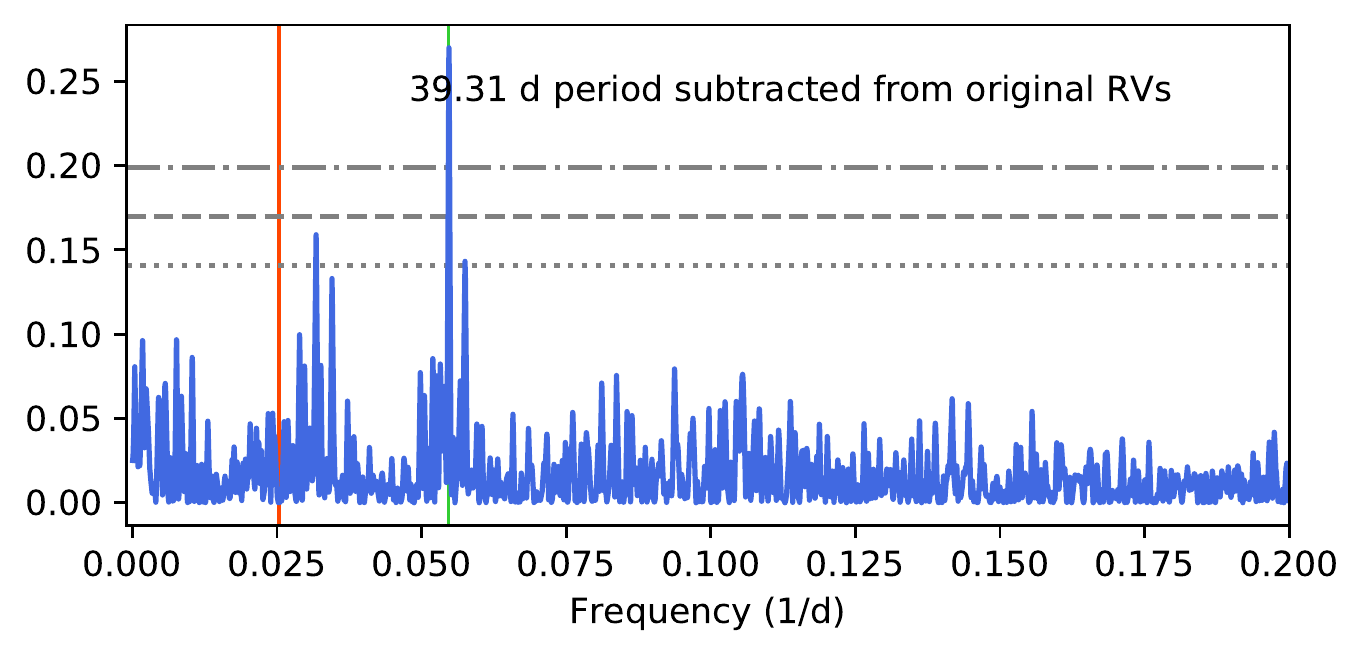}
\caption{GLS peridogram when the 39.31 d signal is subtracted from the original
RV measurements. 
Values
corresponding to a FAP of 10\%, 1\%, and 0.1\% are shown with
horizontal grey lines. The vertical red line indicates the period at 39.31 d
while the vertical green line shows the 18.27 d period.
} 
\label{rv_periodogram_39d_signal}
\end{figure}

  We note that if the 39.31 d signal is firstly prewhitened from the raw RV dataset,
  then the 18.27 d signal clearly remains, Figure~\ref{rv_periodogram_39d_signal}. That suggests that both
  signals are not connected.

   For the sake of completeness we also show the GLS periodogram derived using the DRS
   RVs in Figure~\ref{rv_series_periodogram_drs}, finding similar results.

%

\section{Origin of the radial velocity variations}\label{origin} 
%

\subsection{Activity indexes}
%
 In order to disentangle the effects of activity in the measured RVs from
 other possible effects we made use of several spectroscopic indicators
 of chromospheric activity, the Ca~{\sc ii} H \& K (S-index), 
 H$\alpha$, and Na~{\sc i} D$_{\rm 1}$, D$_{\rm 2}$ lines.
 Although these quantities are provided 
 by the TERRA pipeline, the
 H$\alpha$, and Na~{\sc i} activity indexes came without uncertainty measurements.
 Therefore, we decided to measure the activity indexes following the definitions
 provided by \cite{2011A&A...534A..30G,2018JOSS....3..667G}.
 
\begin{figure}[!htb]
\centering
\includegraphics[scale=0.65]{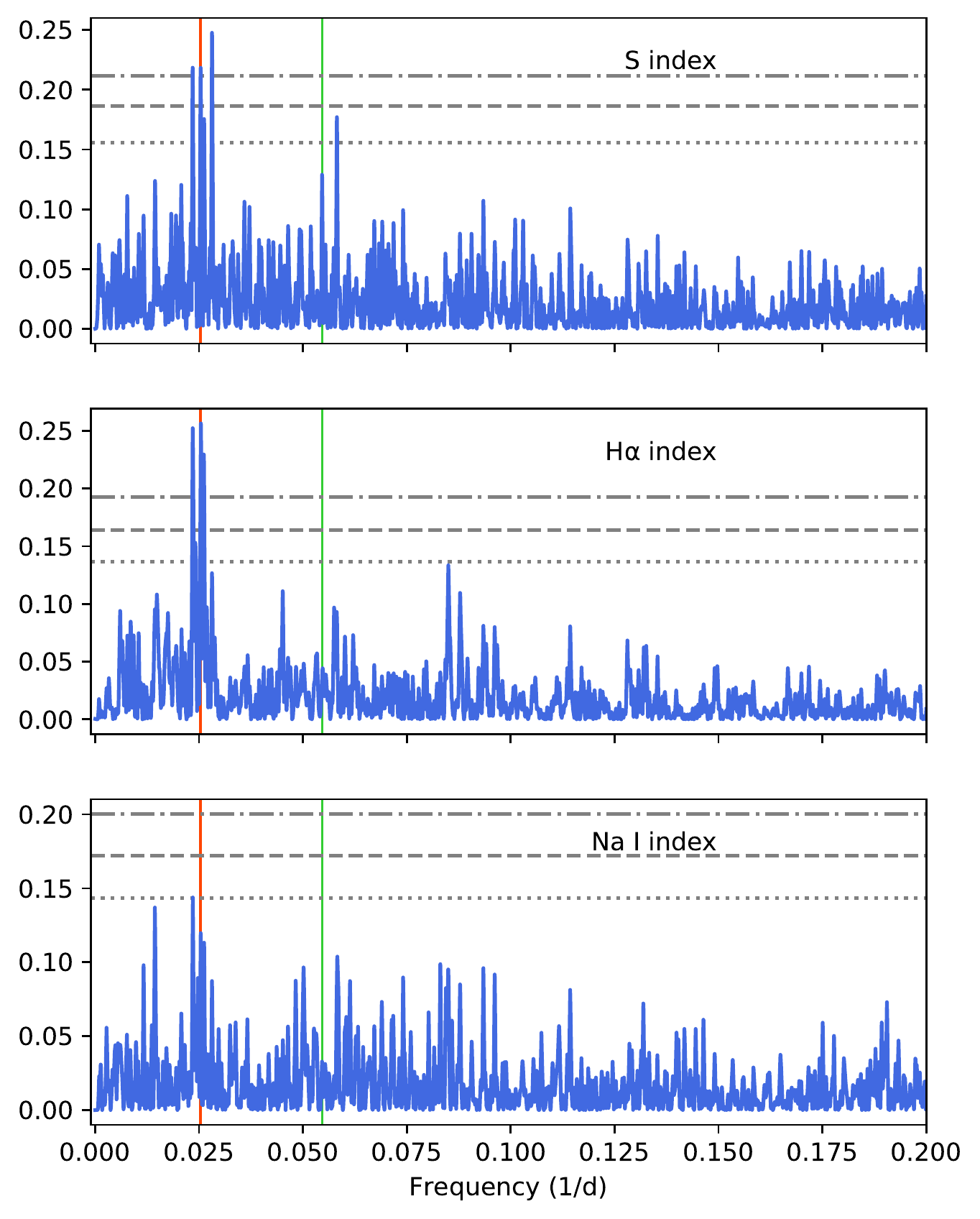}
\caption{
From top to bottom: GLS periodogram of the S, H$\alpha$, and Na~{\sc i} activity indexes
after the subtraction of the long-term activity trend.
The vertical red line indicates the RV period at 39.31 d
while the vertical green line shows the RV period at 18.27 d.  
Values
corresponding to a FAP of 10\%, 1\%, and 0.1\% are shown with
horizontal grey lines.
}
\label{activity_indices}
\end{figure}

 Figure~\ref{rv_series}  shows the temporal variation of 
 the different activity indexes while the periodogram analysis is given in
 Figure~\ref{activity_indices}.
 A long-term activity trend 
 in the three time series is seen. 
 The trend has a period of $\sim$ 34500 d, 20400 d, and 56500 d, in S-index, H$\alpha$, and
 Na~{\sc i} index, respectively, suggesting an activity cycle of more than 55 yrs 
 \footnote{For the Na~{\sc i} index analysis, one observation was excluded
 due to its large uncertainty.}.
 Note that these periods are much longer than the observation timespan, so they are
 extremely uncertain. 
 Once the long-term activity trend is subtracted by a sinusoidal fit,
 a group of peaks with periods in the range 35-45 d is found in all the activity indexes.
 The highest peaks are located at 35.55 $\pm$ 0.05 d (S-index), 39.27 $\pm$ 0.03 d (H$\alpha$-index),
 and 42.54 $\pm$ 0.10 d (Na~{\sc i}-index). 
 A clear peak close to the RV signal at 39.31 d is found in the three indexes. 

 Our analysis also shows that there is a peak close to 18.27 d in the S-index although
 it is not significant. 
 No significant peaks seem to be present at 18.27 days
 neither in the H$\alpha$-index,
 nor in the Na~{\sc i}-index. 

\begin{figure}[!htb]
\centering
\includegraphics[scale=0.50]{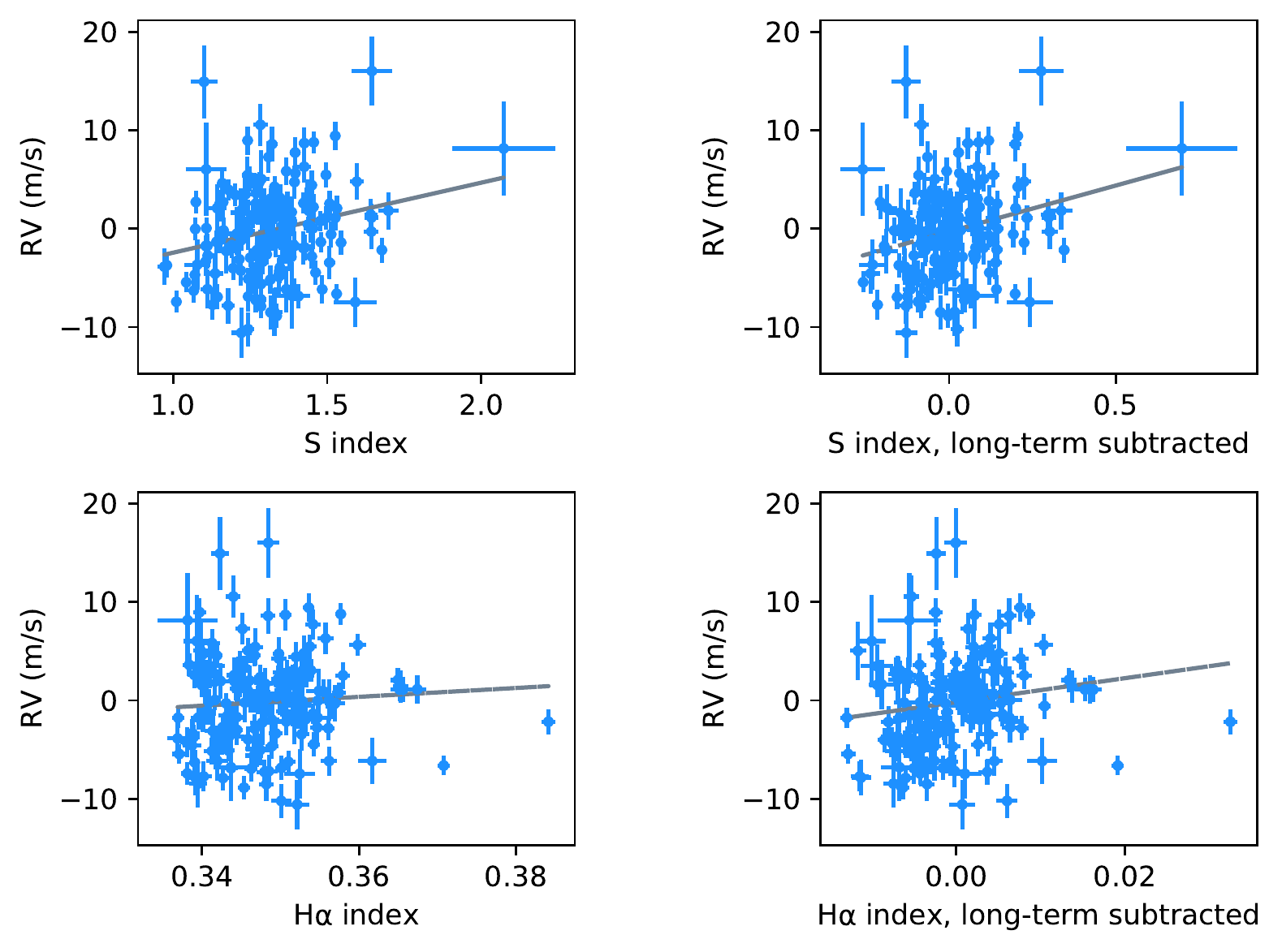}
\caption{
Activity indexes vs. radial velocities. Top left: S-index. 
Top right: S-index after the subtraction of the long-term activity trend.
Bottom left: H$\alpha$-index.
Bottom right: H$\alpha$-index after the subtraction of the long-term activity trend.
The grey line shows the best linear fit. 
}
\label{activity_indices_rv}
\end{figure}

 As an additional test we checked whether our derived RVs show any correlation
 with the activity indexes 
 finding no significant correlation between these
 quantities. The corresponding plots are shown in Fig.~\ref{activity_indices_rv}.
 For the S-index the Spearman's rank $\rho$ is  0.1890 $\pm$  0.0752
 with a z-score = 1.058  $\pm$ 0.431 while for RVs and H$\alpha$-index we obtain
 $\rho$ = 0.1214 $\pm$ 0.0772 and z-score = 0.675 $\pm$  0.433.
 If the activity indexes are corrected by the long-term activity trend, we then obtain
 $\rho$ =  0.2384 $\pm$  0.0756 and z-score = 1.346  $\pm$ 0.4444 for the S-index,
 and $\rho$ = 0.2283 $\pm$  0.0756 and z-score = 1.286  $\pm$ 0.441 for the  H$\alpha$-index. 
 The statistical tests were performed
 by a bootstrap Monte Carlo simulation plus a Gaussian random shift of each data point
 within its error bars 
 \citep{2014arXiv1411.3816C}\footnote{https://github.com/PACurran/MCSpearman/}.

Finally, we also computed 
  the autocorrelation function (ACF) for the RVs, S-index, and H$\alpha$ indexes.
  The corresponding plots are shown in Fig.~\ref{acf_functions} (left). 
   The ACF function has been computed following the prescriptions given by \cite{1988ApJ...333..646E}, 
    using a python wrapper developed by \cite{2015MNRAS.453.3455R}. 
    It is clear that the three datasets show a periodicity around $\sim$ 40 d. But the plot also shows that only in the RV data there is another periodicity at $\sim$ 20 d. In order to confirm that, we also computed the GLS periodogram of the ACFs,
 Fig.~\ref{acf_functions} (right).
   While in the RV dataset the periods at $\sim$ 18.27 and 39.31 d are clearly visible, the ACF function of the activity indicators do show only the period at 39.31 d.
 We note that the ACF of the H$\alpha$ index shows a peak around 17.3 d, but it is not statistically significant (indeed, there is
 a more significant period  at $\sim$ 11.8 d).

\subsection{Periodogram power as a function of time}

Another way to disentangle RVs variations due to the presence of
 planets from stellar activity is to study the evolution
 of the periodogram power of the RVs periods as a function of the number of
 observations \citep[e.g.][]{2016A&A...593A.117A,2017A&A...601A.110M}. 
 Activity regions change in shape and position with time, thus
 producing incoherent (in amplitude and phase) signals. In the
 periodograms this incoherency translates into wider peaks and/or
 a bunch of peaks aside the central one. On the other hand, a keplerian
 signal gains in power with time, thanks to its coherency.

\begin{figure*}[!htb]
\centering
\begin{minipage}{0.48\linewidth}
\includegraphics[scale=0.375,angle=270]{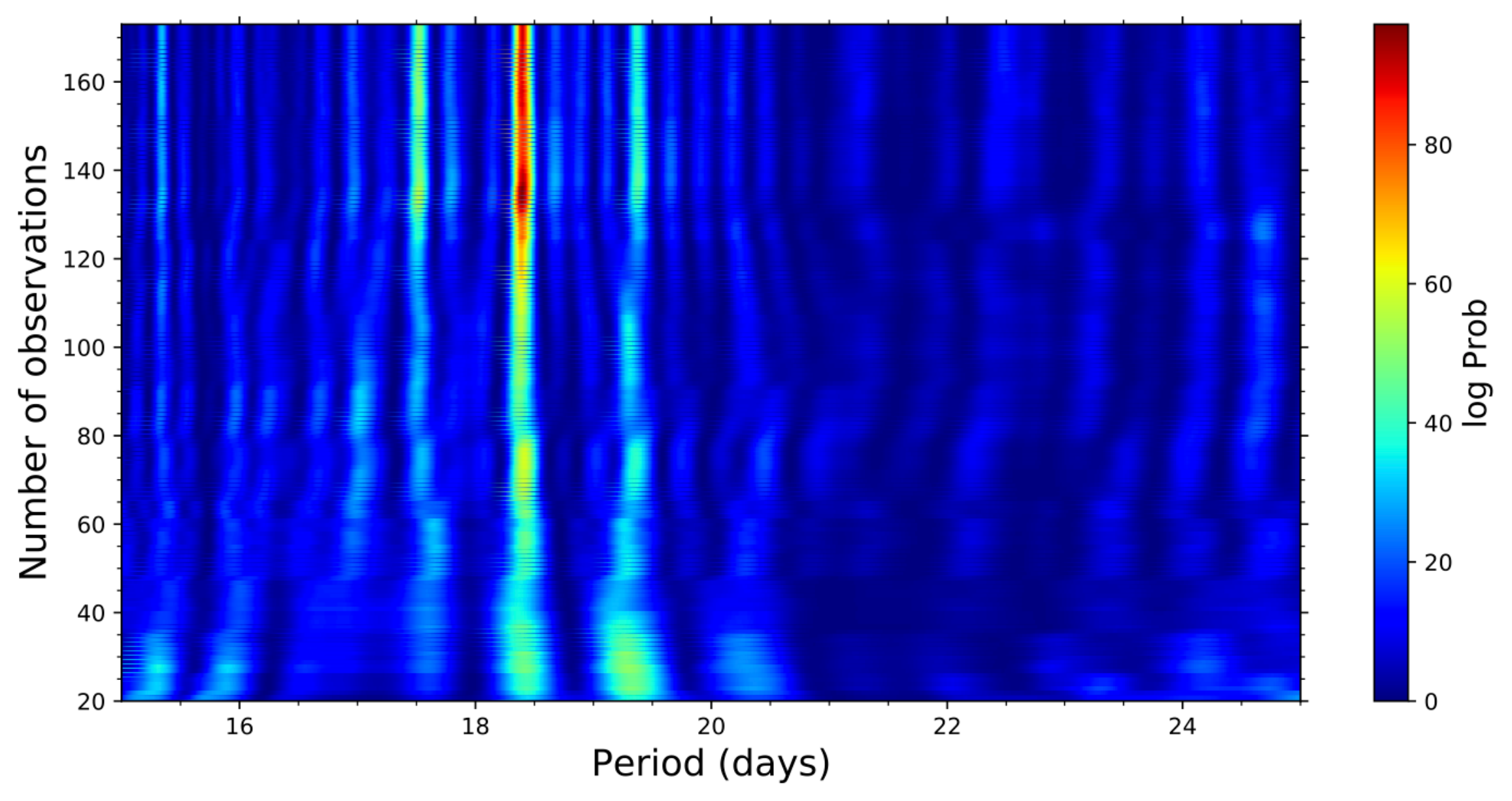}
\end{minipage}
\begin{minipage}{0.48\linewidth}
\includegraphics[scale=0.375,angle=270]{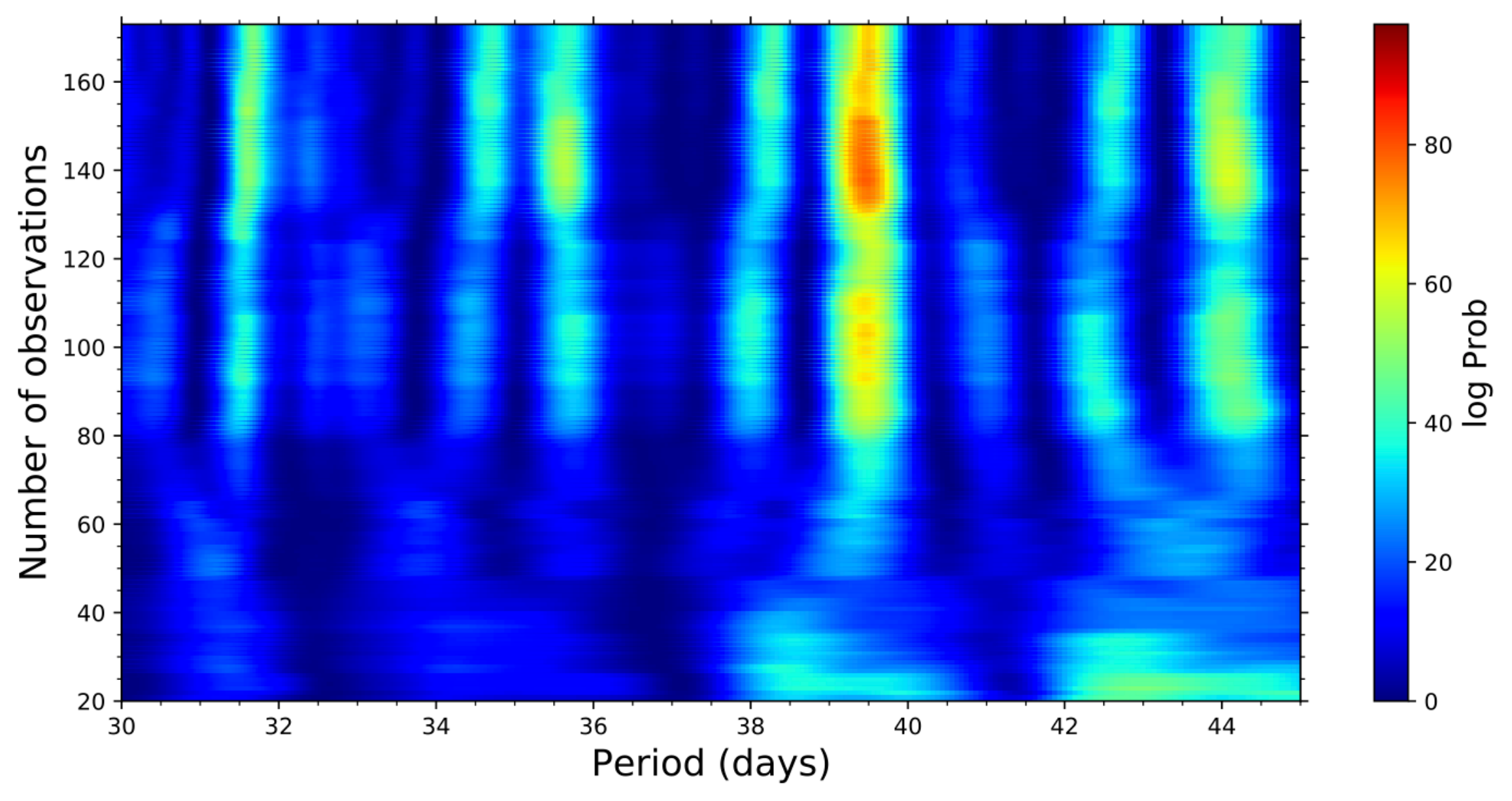}
\end{minipage}
\caption{
Stacked-BGLS periodogram of the HARPS-N RV data of GJ 9689.
Left: zoom around the 15 - 25 d region. Right: zoom around the 30 - 45 d region.}
\label{power_vs_time_activity}
\end{figure*}

 Figure~\ref{power_vs_time_activity} shows the variation of the periodogram
 power as a function of the number of observations for the
 periods found in the RVs time series. 
 For this exercise we use the Bayesian generalised Lomb-Scargle periodogram
 \citep[BGLS,][]{2015A&A...573A.101M}
 which computes the relative probability between peaks.
 
  The analysis 
 reveals that even at a relatively low number of
 observations, around 35-40,
 a period between 18 and 19 d is visible. At around $\sim$ 110 observations the period is well stabilised
 at $\sim$ 18.27 d and since then, it remains constant in period and $\log$Prb
 (Fig.~\ref{power_vs_time_activity}, left).

 The analysis also reveals a signal at a period slightly
 larger than 40 d (although at a rather modest probability),
 however, this signal disappears
 at around 30 observations.
 It is likely the $\sim$ 39 d signal but not well constrained due to the low number
 of observations. 
 A new period around 39 d
 appears again when the number
 of observations is around 80. The period of this signal is not well-constrained
 and it is worth noticing that it is accompanied by many other signals
 (Fig.~\ref{power_vs_time_activity}, right).
  
%

\subsection{Photometry}

\subsubsection{EXORAP photometry} 

\begin{figure}[!htb]
\centering
\includegraphics[scale=0.55]{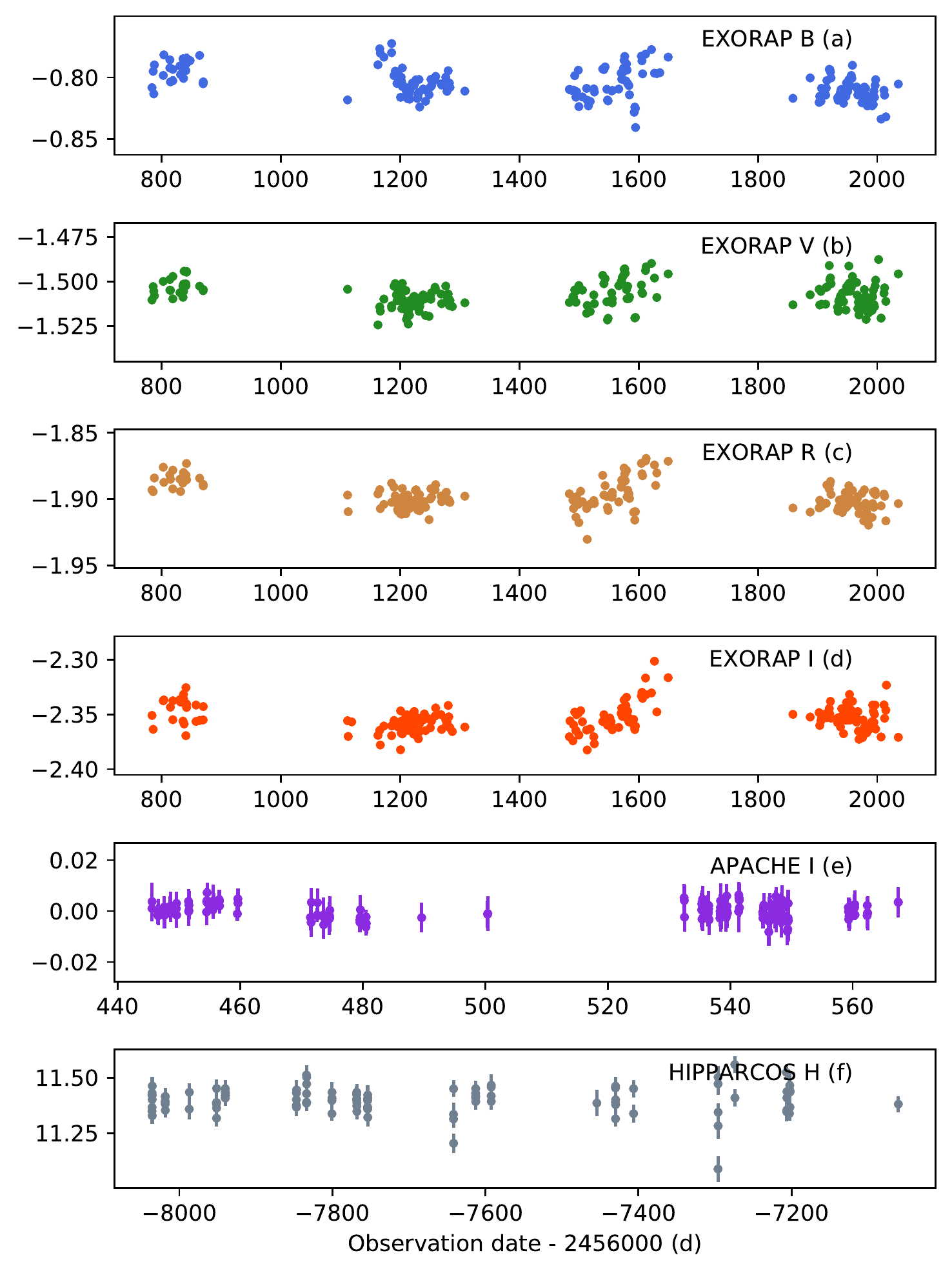}
\caption{Original photometry time series: (a) EXORAP B, (b) EXORAP V, (c) EXORAP R,
(d) EXORAP I, (e) APACHE I, (f) {\sc Hipparcos} H. 
Note that the observations are not contemporaneous.
For EXORAP observation dates are in JD, for APACHE in HJD, while for {\sc Hipparcos} observation dates are in BJD.
} 
\label{fotometry_all_series}
\end{figure}

GJ 9689 was observed photometrically within the EXORAP
 (EXOplanetary systems Robotic APT2 Photometry) 
 program from MJD = 56783 d to MJD = 58034 d.
 Observations were carried out at the Serra la Nave observatory
  on Mt. Etna (Italy) using a 80-cm f/8 Ritchey-Chretien robotic
   telescope (Automated Photoelectric Telescope, APT2).
 A total of 203, 201, 202, and 209 observations were obtained in the
 photometric bands B, V, R, and I, respectively.
 Figure~\ref{fotometry_all_series}, panels (a) to (d), shows the corresponding
 light curves.

 A search for periodicities reveals the presence of a long term signal
 with periods of 
 $\sim$ 244 d (B), 
 $\sim$ 256 d (R),
 $\sim$ 270 d (I).
 No long-term trend is found in band V.
 Once these long-term trends are subtracted, the corresponding GLS analysis,
 Figure~\ref{fotometry_all_period} panels (a)-(d), reveals the presence
 of several signals with periods between $\sim$ 35 and 39 days.
 The highest periods are found at 
 34.87 $\pm$ 0.15 d (B),  38.51 $\pm$ 0.16 d (V), 38.85 $\pm$ 0.16 d (R) and 34.97 $\pm$ 0.18 d (I).
 The significance of these peaks is
 better than 0.1\% in bands B, V, and R, but slightly lower than 10\% in band I.  
%
 The periodograms of the V and R bands (Figure~\ref{fotometry_all_period} panels (b) and (c))
 also show some amount of power in the region around 18 days.
 However, no clear peaks are found at 18.27 days.
%

 A comparison of Fig.~\ref{rv_series}  and Fig.~\ref{fotometry_all_series} shows that 
 the star is brighter when the activity S-index is lower.
 This behaviour is quite different from what we observe in the Sun (and similar stars)
 where long-term variability is dominated by faculae.
 We thus speculate that the stellar activity in GJ 9689 is spot-dominated
 \citep{2018ApJ...855...75R}.

\subsubsection{APACHE photometry} 
%
%
 The APACHE (A PAthway toward the Characterisation of Habitable Earths)
 photometric survey \citep{2013EPJWC..4703006S} observed GJ 9689 with
 a 40-cm telescope located in the Astronomical Observatory of the
 Autonomous Region of the Aosta Valley. The observations cover a 
 time span of 122 days from  HJD = 2456445 d to  HJD = 2456567 d. The number of observations
 amounts to 158. A Johnson I filter was used to carry out the observations.
 A clear significant period is found at 38.22 $\pm$ 1.04 d in the GLS analysis.

Note that APACHE and EXORAP I datasets have different sampling, baselines, and data quality
 so it is not surprising to find slightly different periodograms. 
 In particular the EXORAP dataset covers a baseline of $\sim$ 1250 d, allowing us the
 detection of a long-term signal ($\sim$ 270 d) while the APACHE baseline is only 122 d.

%
\subsubsection{Hipparcos H band photometry} 
\begin{figure}[!htb]
\centering
\includegraphics[scale=0.65]{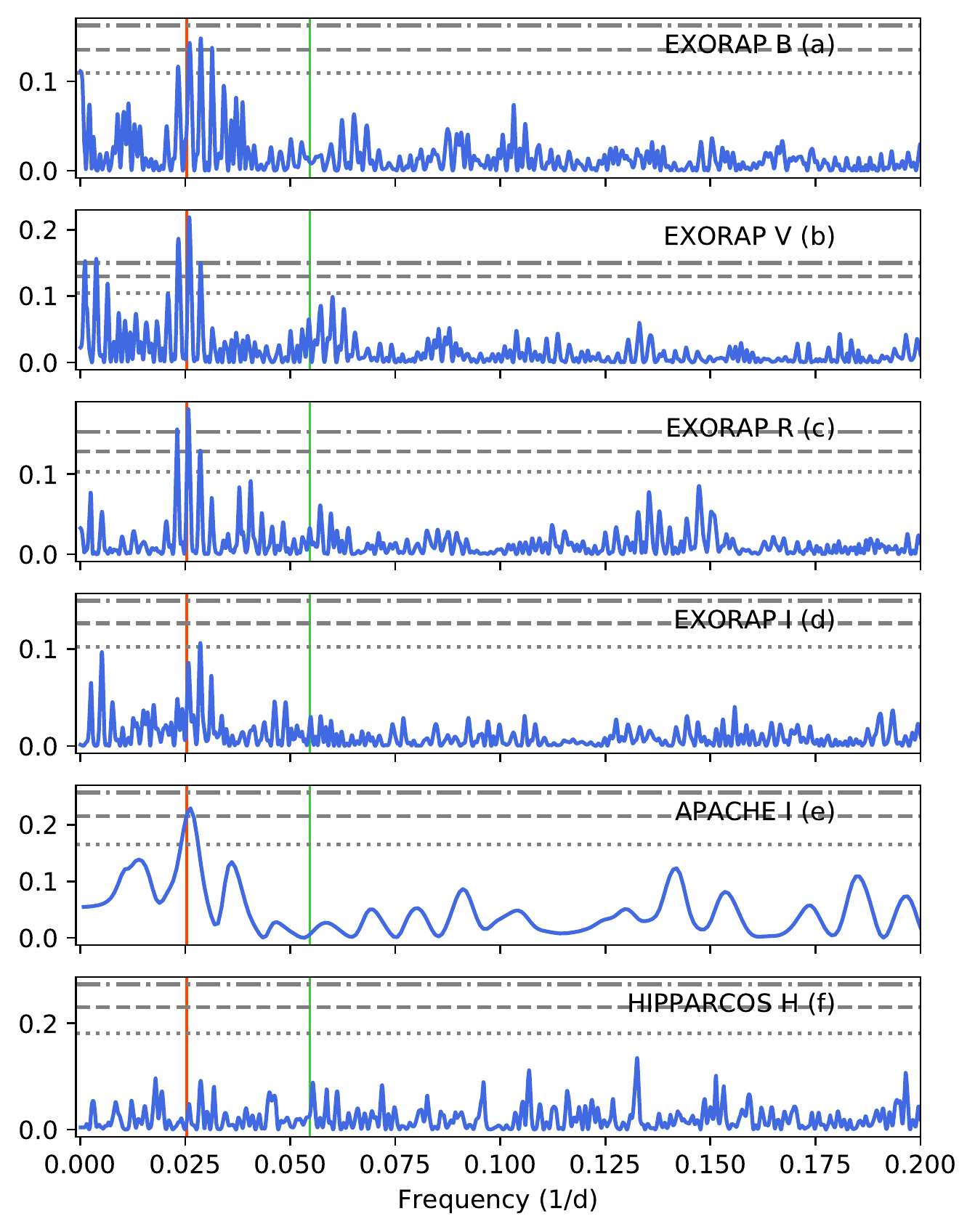}
\caption{GLS periodograms of the photometry data sets. Panels (a) , (c), and (d)
show the EXORAP B, R, and I band photometry after subtracting the corresponding long term periods (see text for details).
Panels (b), (e), and (f) show the original (i.e. no long term period subtracted) EXORAP V,
APACHE I and {\sc Hipparcos} H photometry analysis. 
Values corresponding to a FAP of 10\%, 1\%, and 0.1\% are shown with
horizontal grey lines.
The vertical red line indicates the period at 39.31 d
while the vertical green line shows the 18.27 d period.} 
\label{fotometry_all_period}
\end{figure}

%
 We also analysed the available {\sc Hipparcos} \citep{1997ESASP1200.....E} H band photometry.
 GJ 9689 was observed during 975 days between BJD = 2447964 d and
 BJD = 2448939 d,  with a total of 101 data points. 
 The corresponding periodogram is shown in panel (f) of Figure~\ref{fotometry_all_period},
 where no significant periods are found. 
 While it is true that a peak seems to be present at $\sim$ 18 d, 
 it is very far from being significant. Furthermore, this region of the periodogram
 is largely crowded with numerous peaks with a similar (or larger) power. 
%
%

%
\subsection{Wavelength dependence}
\begin{figure}[!htb]
\centering
\includegraphics[scale=0.65]{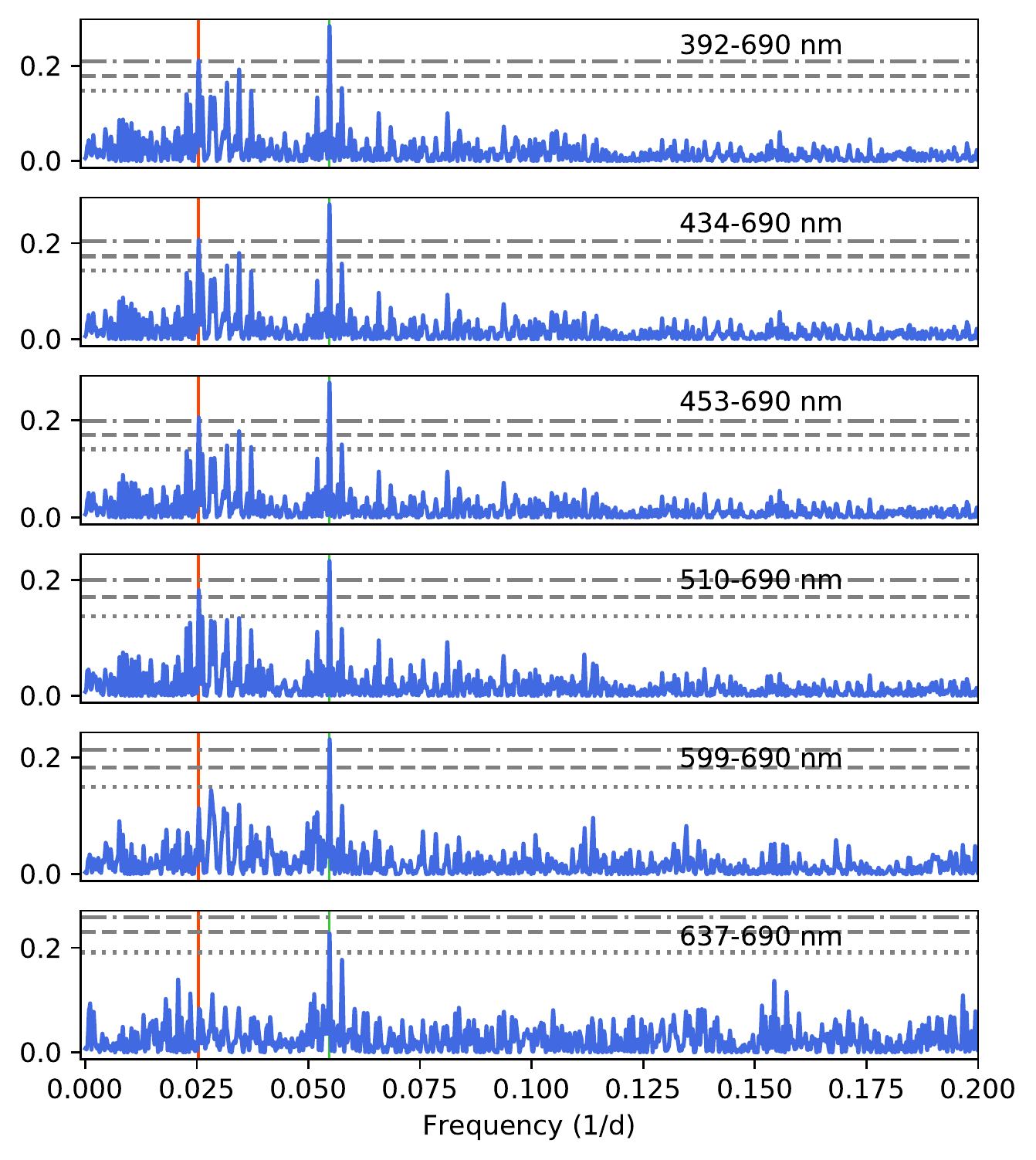} 
\caption{GLS periodograms of the TERRA RVs as a function of the blue cut-off wavelength.
 From top to bottom: RVs are computed using the 1, 16, 22, 37, 55, and 61   
 aperture as blue cut-off. 
 Values corresponding to a FAP of 10\%, 1\%, and 0.1\% are shown with
 horizontal grey lines.
 The vertical red line indicates the period at 39.31 d
 while the  green line shows the 18.27 d period.} 
\label{gls_terra_orders}
\end{figure}

%
 We also explored the dependence of the periodic signals identified
 in the RV analysis on the wavelength. 
 In order to do that we proceeded as in \cite{2013A&A...549A..48T}
 and exploited the fact
 that the TERRA pipeline allows us to select a
 blue cut-off aperture when computing the RVs.
 The results are shown in Figure~\ref{gls_terra_orders} that shows
 the GLS periodogram of the RVs obtained by using different
 blue cut-off wavelengths. It can be seen that the period at 
 18.27 days is always visible, independently of the bluest \'echelle
 aperture used in the RVs computations.
 On the other hand, the period at 39.31 days disappears when
 only wavelengths redder than $\sim$ 600 nm \space are used in the
 RVs computation (two bottom panels).
%

%
%
%
\subsection{CCF Asymmetry diagnostics}
\begin{figure}[!htb]
\centering
\begin{minipage}{0.48\linewidth}
\includegraphics[scale=0.30]{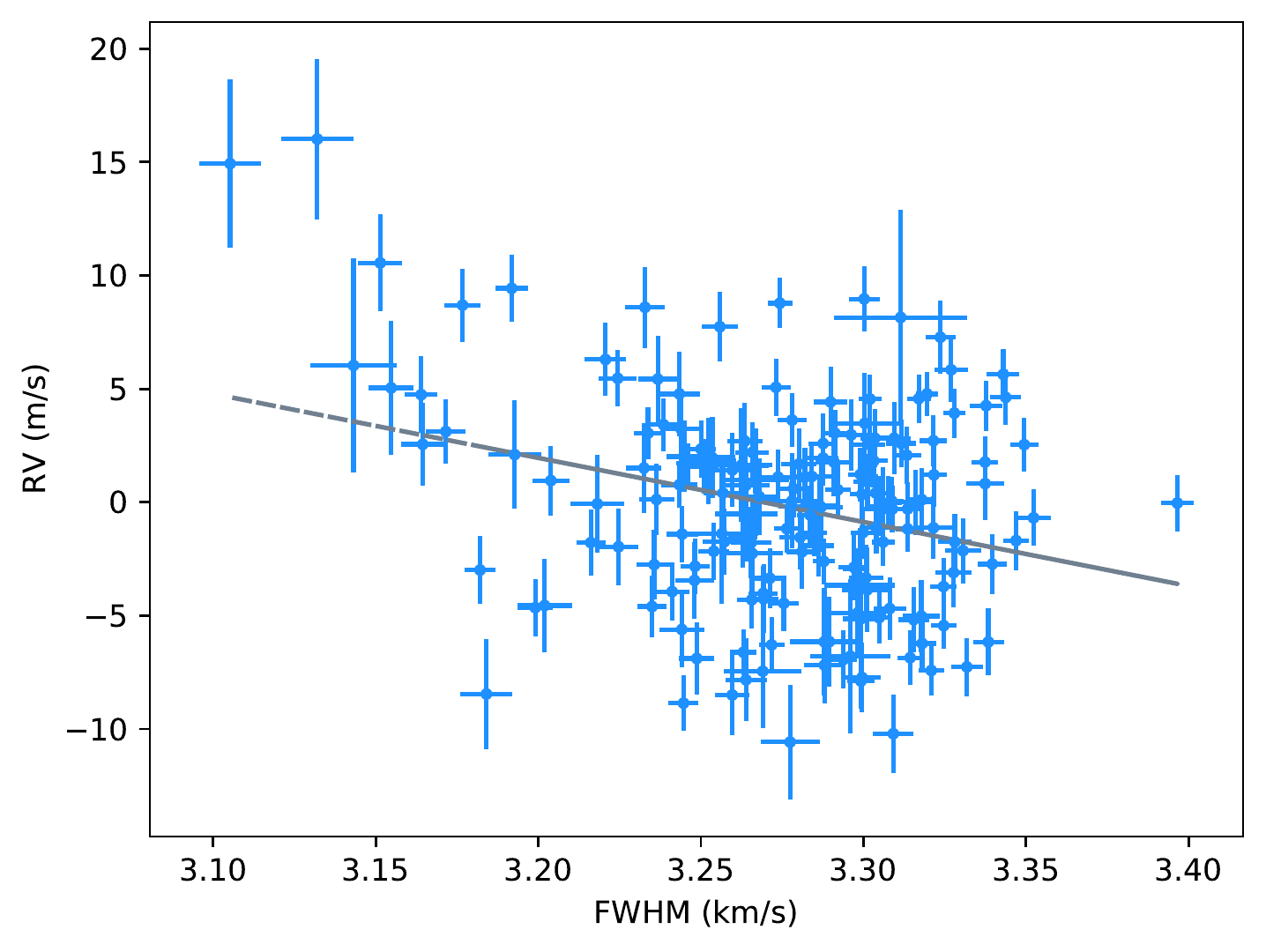}
\end{minipage}
\begin{minipage}{0.48\linewidth}
\includegraphics[scale=0.30]{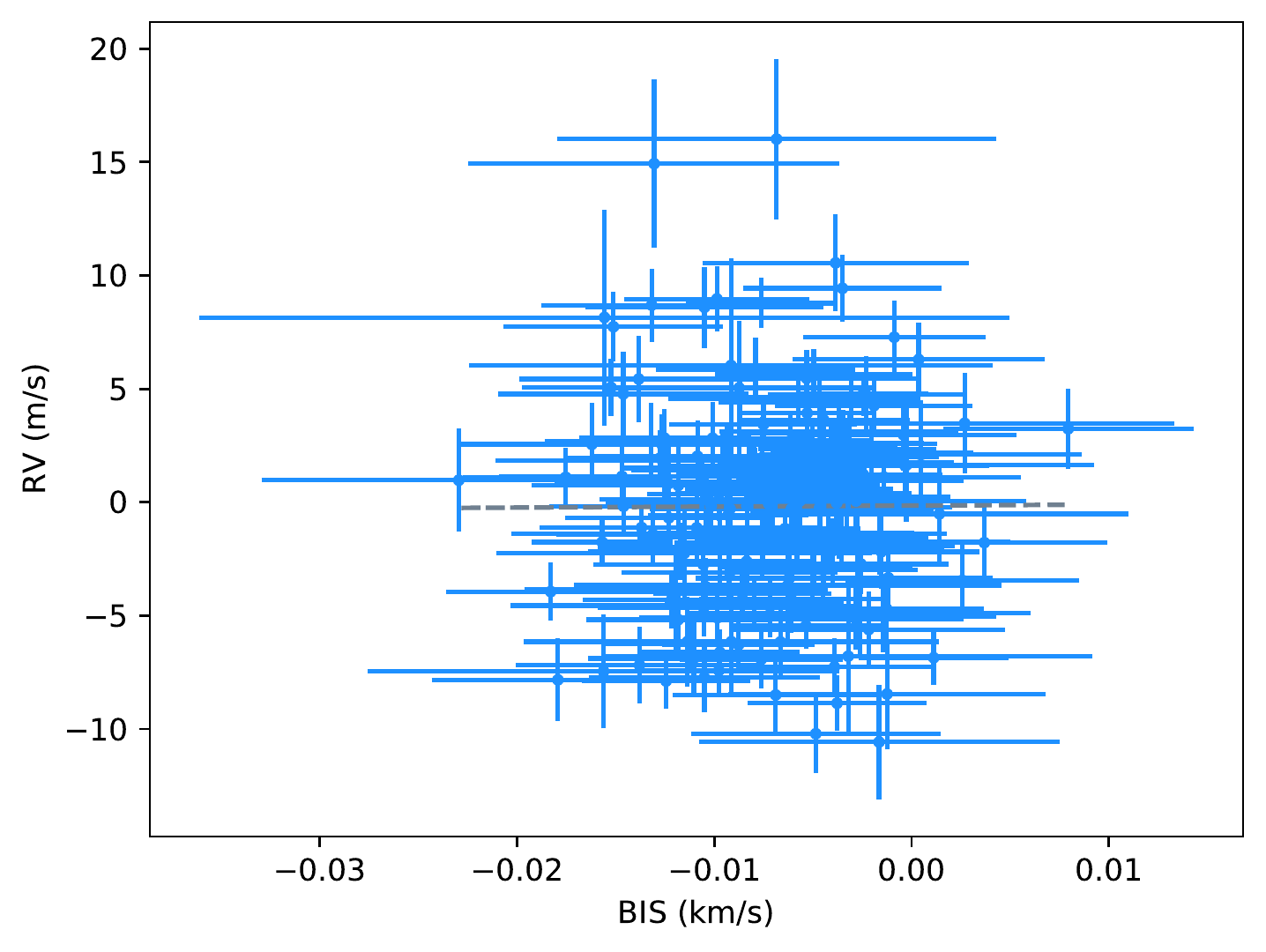}
\end{minipage}
\begin{minipage}{0.48\linewidth}
\includegraphics[scale=0.30]{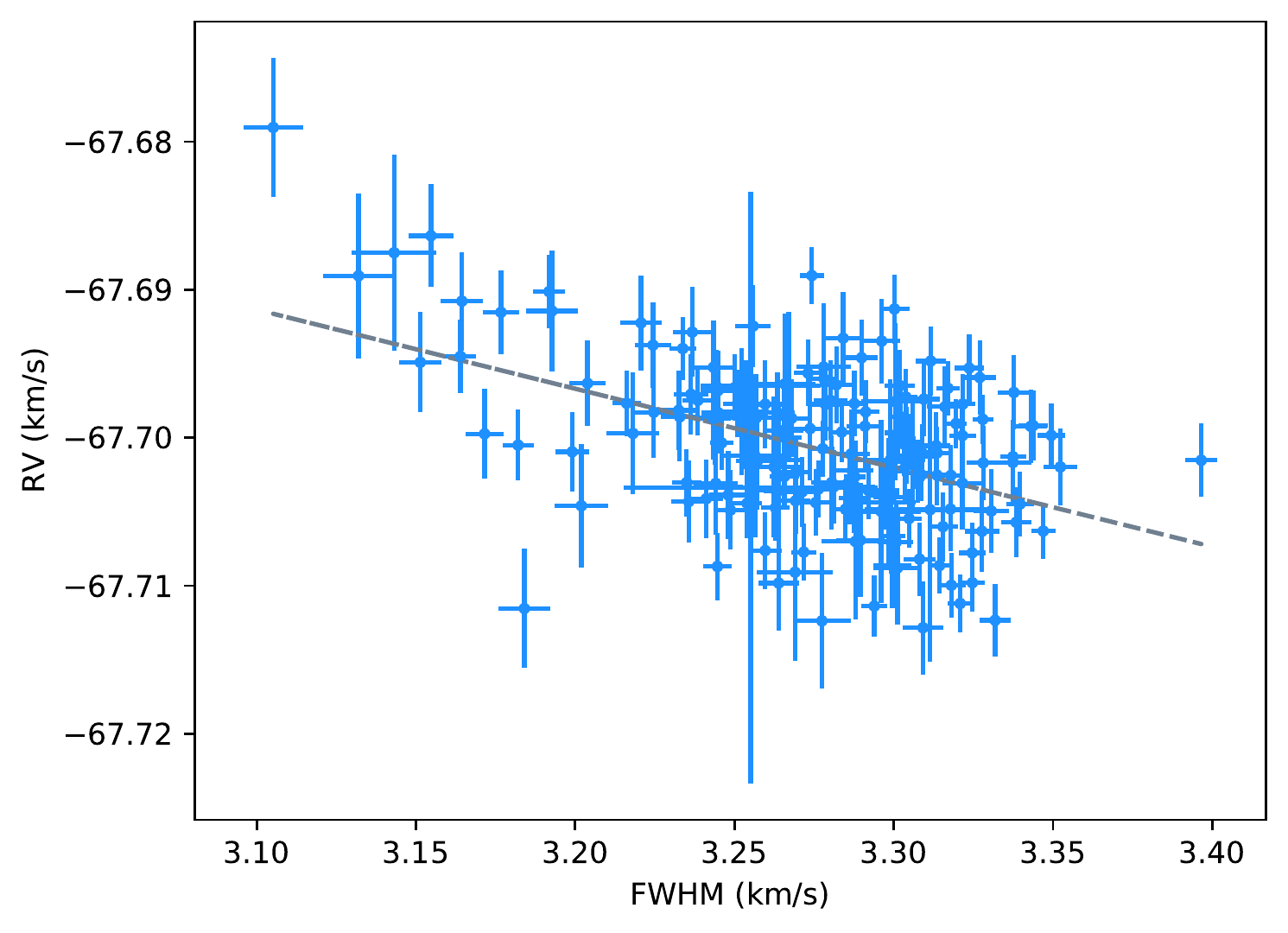}
\end{minipage}
\begin{minipage}{0.48\linewidth}
\includegraphics[scale=0.30]{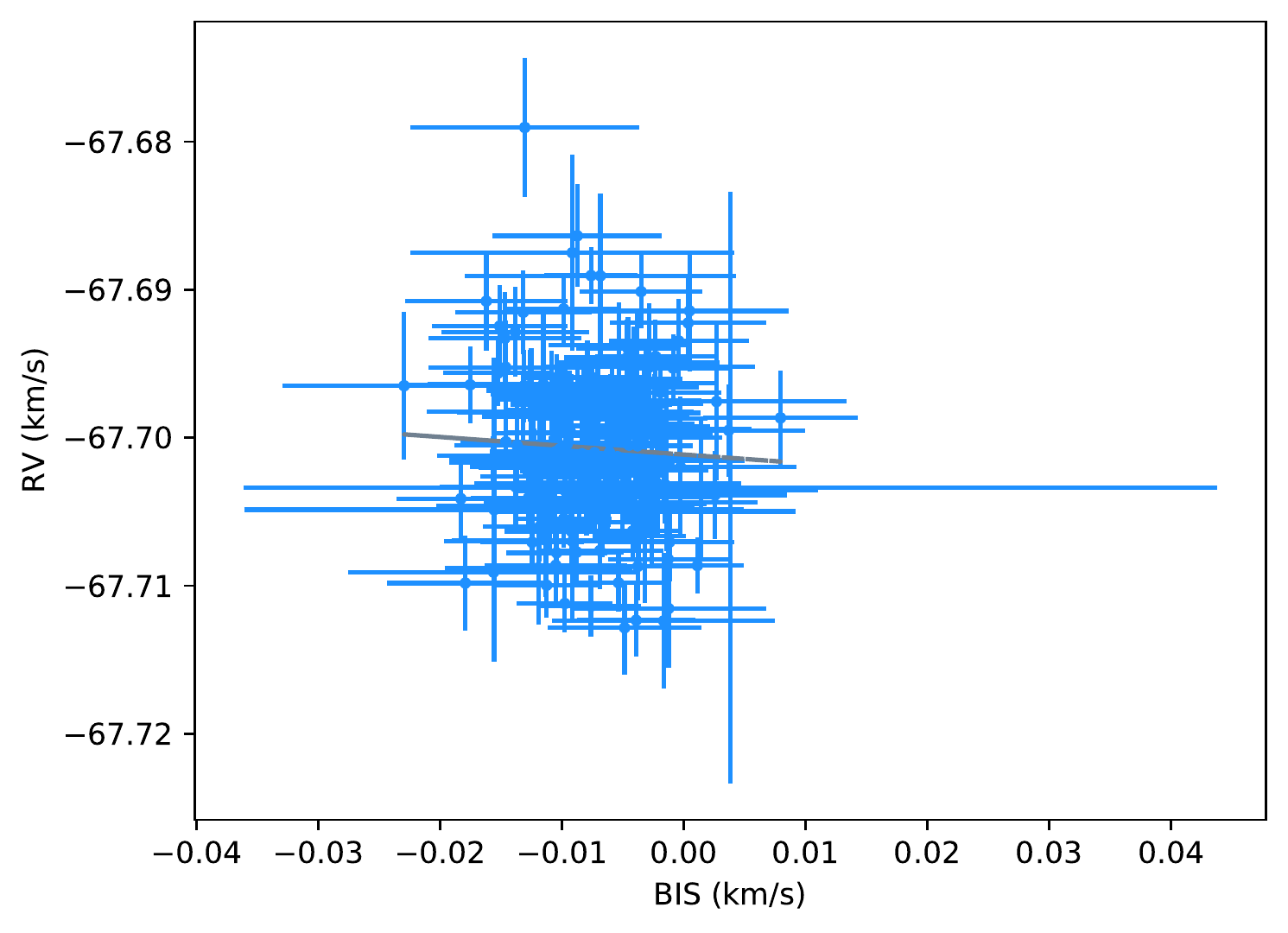}
\end{minipage}
\caption{ CCF asymmetry diagnostics vs. radial velocity. Left, FWHM. Right, BIS.
The grey line shows the best linear fit.
Upper panels show the TERRA RVs, while bottom panels show the results for the DRS
RVs.
}
\label{asymetry_indices}
\end{figure}

%
 Stellar spots and pulsations are known to affect the 
 shapes and the centroids of the spectral lines.
 We therefore investigated for
 possible correlations between the RVs and several CCF asymmetry diagnostics,
 namely the CCF width (FWHM) and its bisector velocity span (BIS)
 \citep[e.g.][]{2001A&A...379..279Q,2009A&A...495..959B,2009A&A...506..303Q}.
 Both quantities are directly provided by the HARPS-N DRS.
 As mentioned before, these quantities should be taken with caution when
 dealing with low-mass stars. 
%
 Figure~\ref{asymetry_indices}  (upper panels) shows 
 the TERRA RVs as a function of the FWHM and BIS values.
 A Spearman's correlation test confirms that there is no monotone dependence
 between the RVs and the measured FWHM values
 ($\rho$ = -0.1304 $\pm$ 0.0807 and z-score = -0.726 $\pm$  0.454)
 or between the RVs and the BIS
 ($\rho$ = 0.0315 $\pm$ 0.0781 and z-score = 0.174 $\pm$ 0.433).
 Furthermore, no significant signals were found in the periodogram analysis
 of the BIS. For the FWHM, a rather broad but statically significant period is found at $\sim$ 402 $\pm$ 8 d,
 but no significant periods remain after this period is prewhitened.
 No signals are found either in BIS or in FWHM around the periods identified in the RV analysis. 

 Given that TERRA RVs are derived using a different method (and a different pipeline),
 a comparison between the FWHM and BIS values with the RVs derived by the CCF performed by
 the HARPS-N DRS is also mandatory.
 The corresponding plots are shown in Figure~\ref{asymetry_indices} (bottom panels).
 We find that the results are similar to the ones obtained using the TERRA RVs, that is,
 there is no dependence between the DRS RVs and the measured FWHM or BIS values
 ($\rho$ = -0.2990 $\pm$ 0.0732 and z-score = -1.712 $\pm$  0.446 for the FWHM, and
  $\rho$ = -0.0087 $\pm$ 0.0797 and z-score = -0.048 $\pm$  0.443 for the case of the BIS).

%
\subsection{Analysis of individual seasons}
%
\label{subsec:seasons}
 In this section we analyse the data by considering three different observing seasons.
 They are indicated with vertical lines in Fig.~\ref{rv_series}.
 Note that 'season 1' and 'season 2' do indeed contain data from two observing seasons.
 This choice was made in order to have enough RV data points in each season. 
 The first season includes 46 data points taken from May 26, 2013, to October 22, 2014.
 The second season covers the observations performed between  August 23, 2015 and
 November 28, 2016 with 62 observations. Finally, we consider 47 observations between
 April 16, and October 10, 2017. The remaining observations are not considered as they
 are sparse in time and amount only up to 19.
 
\begin{figure*}[!htb]
\centering
\begin{minipage}{0.48\linewidth}
\includegraphics[scale=0.65]{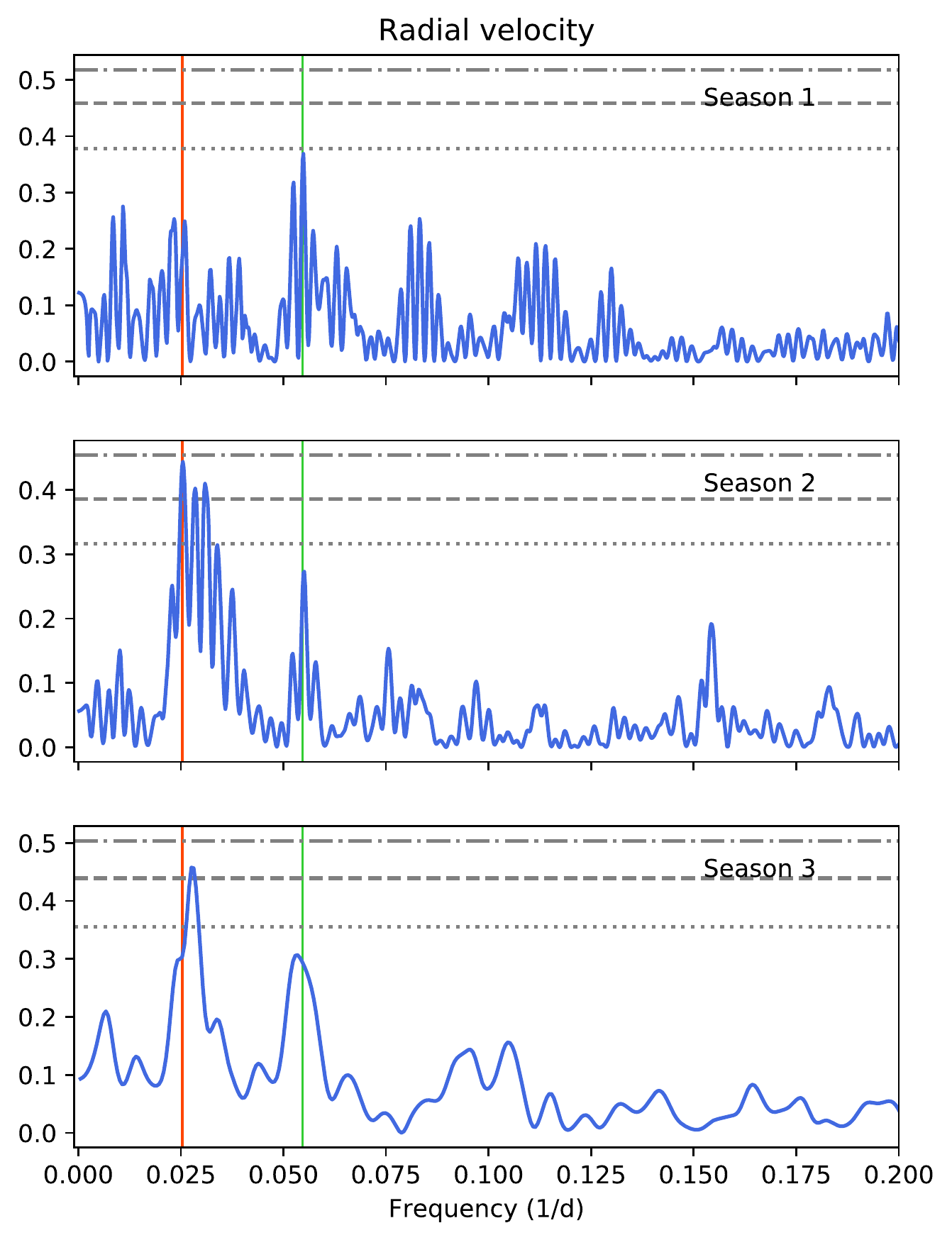}
\end{minipage}
\begin{minipage}{0.48\linewidth}
\includegraphics[scale=0.65]{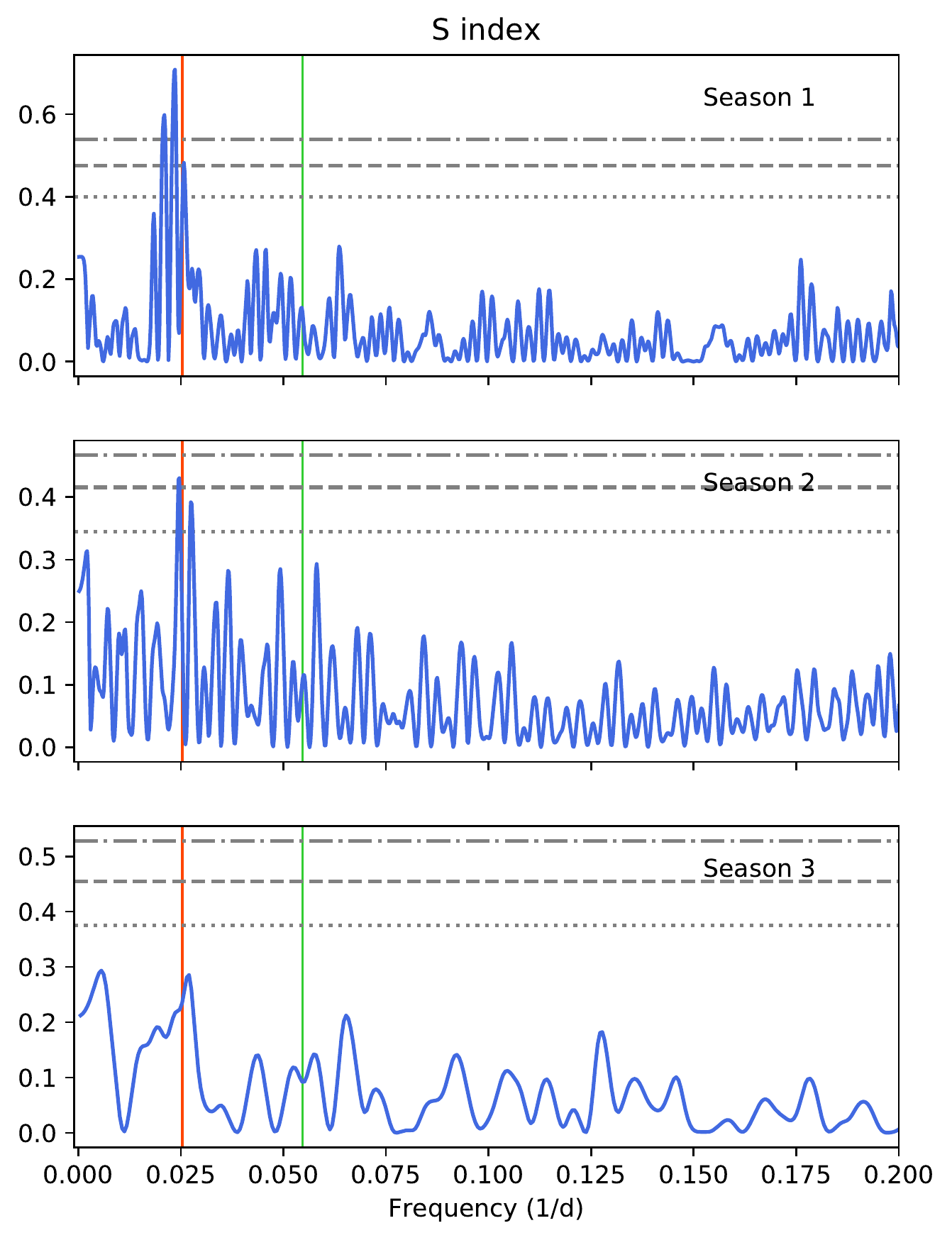}
\end{minipage}
\caption{GLS periodograms for the different seasons analysed.
Left: RVs. Right: S-index. 
Values corresponding to a FAP of 10\%, 1\%, and 0.1\% are shown with
horizontal grey lines. The vertical red line indicates the period at 39.31 d
while the vertical green line shows the 18.27 d period.} 
\label{seasons}
\end{figure*}

 The RV and S-index periodogram for each
 season are shown in Fig.~\ref{seasons} (left, and right, respectively).
 The vertical red and  green line indicate the identified periods at 39.31 d
 and 18.27 d, respectively.
 The figure clearly shows that the RV period at 18.27 d is visible in all seasons, 
 with a similar structure and power. However, the structure of the RV period at 
 39.31 d changes from one season to another. The peak is located at 38.53 d,
 39.25 d, and 35.93 d in the first, second, and third season, respectively.
 Its power also changes with time, being the dominant signal in the second and
 third seasons. We note that in these seasons, the values of the S-index are higher than in 
 the first season (see Fig.~\ref{rv_series}), supporting the hypothesis that the $\sim$ 39.31 d period is due to stellar activity. 
 Regarding the S-index (right panel), there are no signals close to 18.27 d in any of the seasons, while
 there are prominent signals close to 39.31 d in all seasons. The structure of these signals
 change from one season to another and the highest periods are located
 at 42.56 d, 40.76 d, and $\sim$ 178 d in the first, and second, and third season, respectively.

\subsection{Spectral window analysis}

 Given that the periods found in the RV time series of GJ 9689 are at
 18.27 d and 39.31 d it is reasonable to ask whether
 the signal at 18.27 d is the first harmonic of the signal
 at 39.31 d.

\begin{figure}[!htb]
\centering
\includegraphics[scale=0.65]{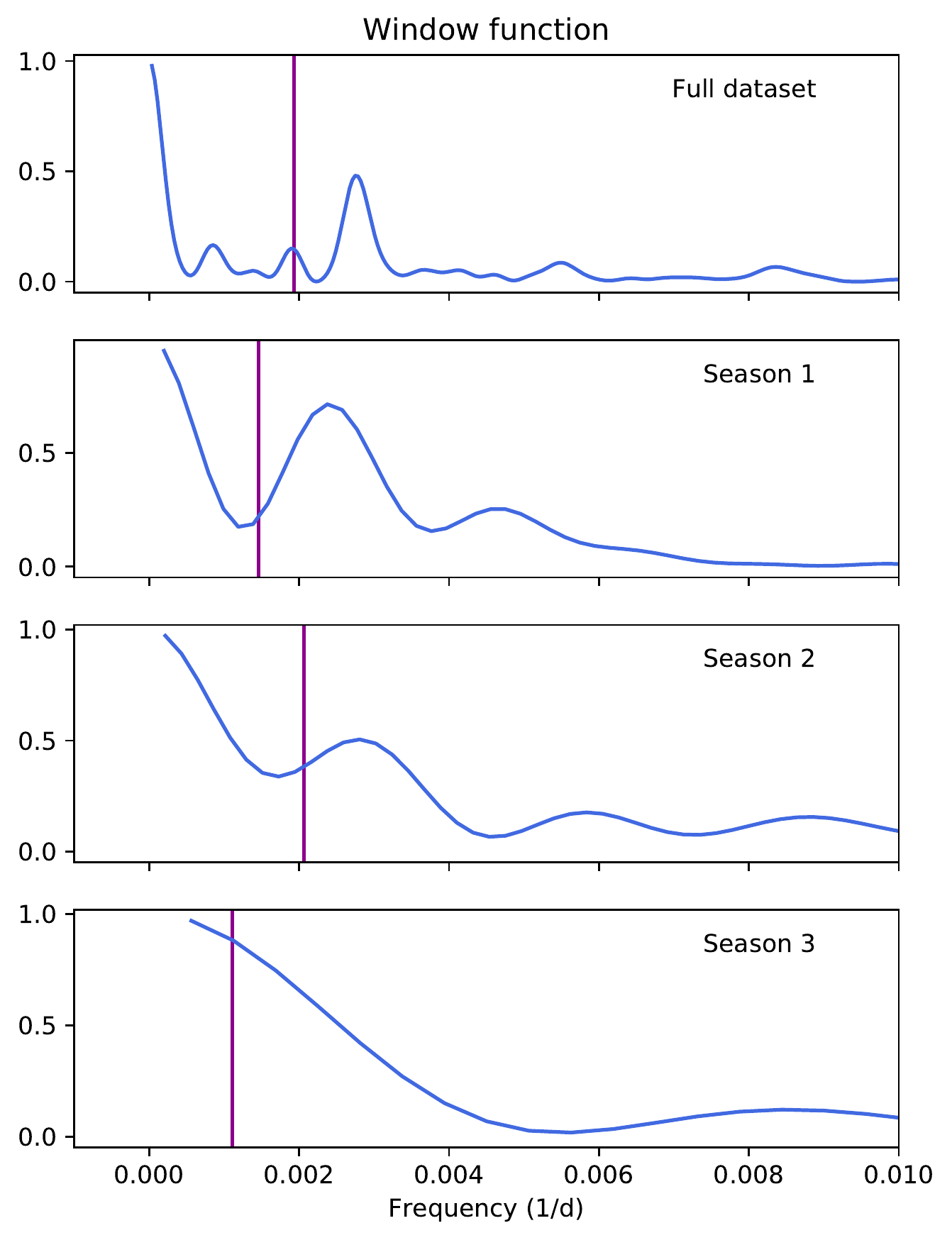}
\caption{Spectral window when the full RV dataset is considered (top) and for each of the different analysed seasons.
The magenta line indicates the frequency that a signal in the spectral window should have in order to produce
an aliasing phenomenon between the $\sim$ 18.27 d and $\sim$ 39.31 d periods.}
\label{M059-window}
\end{figure}

 To answer this question, Figure~\ref{M059-window} (top panel) shows the spectral window of the
 original RV dataset of GJ 9689. There are three prominent peaks
  at frequencies 0.000854 d$^{\rm -1}$, 0.001895 d$^{\rm -1}$,
   and 0.002749 d$^{\rm -1}$.
 These peaks are related to the gaps in the
 RV curve (Fig.~\ref{rv_series}). There is the obvious 1 cycle/year peak (0.00275 d$^{\rm -1}$)
 and two peaks related to the poor sampling in the 2458000 - 2459100 BJD
 interval: 1 cycle/1100 d (0.0009 d$^{\rm -1}$) and 1 cycle/550 d (0.0018 d$^{\rm -1}$).
  The presence of a peak in the spectral window at 0.001895 d$^{\rm -1}$
   might indicate an alias phenomenon as
0.001895 d$^{\rm -1}$ + 0.025436  d$^{\rm -1}$ (frequency of the 39.31 d period) is 0.027331 d$^{\rm -1}$,
	       which is close to 0.054734/2 d$^{\rm -1}$ ($\approx$ 0.027367 d$^{\rm -1}$, i.e., half the frequency of the 18.27 d period).

 To further investigate this possibility, we also analysed the spectral window in the three different seasons considered before.
 The results are shown in Fig.~\ref{M059-window} where for each season we indicate with a magenta line the position that a peak
 in the spectral window should have in order to produce the aliasing phenomenon between the 39.31 d and 18.27 d periods. The analysis
 takes into account that the signals appear at slightly different frequencies in the different seasons. For example, in season one,
 the 39.31 d signal appears at 38.53 d (f = 0.025953 d$^{\rm -1}$) while the 18.27 d signal is at 18.24 d (f = 0.045823 d$^{\rm -1}$). If the 38.53 d and
 18.27 d were related by an alias phenomenon due to a signal in the window function, this signal should appear at a frequency f$_{\rm win}$
 $\approx$ 0.001458 d$^{\rm -1}$. As it can be seen in Fig.~\ref{M059-window} (second panel from top), there is no such a signal in the window function of season one. 
 The same happens in the second season (third panel from top).
 In this case, the 39.31 d signal appears at 39.24 d (f = 0.02548 d$^{\rm -1}$) and the 18.27 d at 18.14 d (f = 0.055122 d$^{\rm -1}$). 
 If the signal at 18.14 d were an alias of the 39.24 d signal, there should be a signal in the window function at a frequency
 f$_{\rm win}$ $\approx$ 0.0021 d$^{\rm -1}$, which is not the case. 
 In the third season (bottom panel) the signals are located at 35.93 d (f = 0.027835 d$^{\rm -1}$), and 18.71 d (f = 0.053457 d$^{\rm -1}$). 
 Again, there are no peaks in the spectral window that might originate an aliasing phenomena. 

 We conclude that it is unlikely that the 18.27 d signal is the harmonic of the 39.31 d, a result which is
 in line with all the different analyses performed before. 

%
\section{Modelling of the radial velocity variations}\label{modelling} 
%
  The analysis performed in the previous section is 
  consistent with a keplerian origin of the coherent signal at
  18.27 d, while the incoherent signal at 39.31 d seems to be related to the
  rotation period of the star.
  This conclusion is based on the following observational facts:

  \begin{itemize}
  \item A periodicity close to 39.31 d is found in the analysis of the main optical activity indicators
  as well as in the analysis of the available photometry.
  \item Its power and frequency changes with the number of observations and from one observing
  season to another. 
  \item This signal tends to disappear if only the reddest region
        of the spectra is used for the computation of the RVs.
  \end{itemize}

  On the other hand, for the 18.27 period:

  \begin{itemize}
    \item It does not seem to be related (to be a harmonic) of the 39.31 days period.
    \item No hint of this period is found in the activity indexes, photometry, or
          CCF asymmetry indicators.
    \item It appears in all observing seasons at a similar frequency and similar power.
          In addition, it does not show variations with the number of observations. 
    \item It is always found in the RVs analysis even if only the reddest region
          of the spectra is considered.
  \end{itemize}

\subsection{Gaussian processes modelling}
%
In order to model the RV data
we have defined a Bayesian framework based on a Monte Carlo sampling of the parameter space
with a Gaussian Processes model (GP).
Before modelling, a 3$\sigma$ clipping algorithm was applied to the RVs to identify potential outliers that might
affect the results. As a consequence, two data points were excluded from the following analysis.
The likelihood function is given by

\begin{equation}
\ln p({y_n},{t_n},{\sigma^2_n},\theta) = -\frac{1}{2} \textbf{r}^T K^{-1} \textbf{r} - \frac{1}{2} \ln{\det K} - \frac{N}{2} \ln{2\pi}
\end{equation}

\noindent where $y_n$, $t_n$, $\sigma_n$ are, respectively, the radial velocities, time of observations and errors, $\theta$ is the array of parameters, $\textbf{r}$ is the residual vector obtained by removing the (deterministic) model from data, $K$ is the covariance matrix and $N$ is the number of observations.

The covariance (or kernel) function adopted in this analysis is a Quasi-Periodic (QP) function and it has been obtained by multiplying an exp-sin-squared kernel to a squared-exponential kernel \citep[{\tt george} python package,][]{hodlr} added to an extra white noise (jitter) term, and it is defined as follows

\begin{equation}
k(i,j) = h^2 \exp{(-\frac{(t_i-t_j)^2}{2 \tau^2})} \exp{(-\frac{\sin^2(\pi(t_i-t_j)/P_{rot})}{2\omega^2})} + \delta_{ij} \sigma_{Jit}
\end{equation}

\noindent where $k(i,j)$ is the $ij$ element of the covariance matrix, $t_i$ and $t_j$ are two times of the RV data set, $h$ is the amplitude of the covariance, $\tau$ is the timescale of the exponential component, $\omega$ is the weight of the periodic component, $P_{rot}$ is the period, $\delta_{ij}$ is the Kronecker delta function and $\sigma_{Jit}$ is the white noise term. 

The parameter space is sampled with {\tt emcee} \citep{2013PASP..125..306F}, based on the affine-invariant ensemble sampler for Markov chain Monte Carlo (MCMC) \citep{goodman2010}.

In this analysis we have compared two models which differ on the presence of a planetary (keplerian) signal
\citep[][]{2018PASP..130d4504F}
but share the effect of a linear trend (characterised by the parameters
$\gamma$, and $\dot{\gamma}$)
, respectively, 'GP-only' and 'Star-planet' models.
 
Priors of the models are reported in Table \ref{tab:priors}, they have been chosen uninformative and as large as possible. This parameter space is covered by 32 walkers, randomly initialised within the priors ranges. This choice on the initial position of the walkers produces very low probability values at the beginning of the emcee chain. These values have been eliminated by a burn-in phase, here set as the first 20K steps. After the burn-in phase a blob, centred at the maximum probability position, is initialised to feed a following chain.
This chain runs until the autocorrelation time of each parameter \citep[see][]{Sokal1996MonteCM}, evaluated every 10K steps, varies less than 1\% and the chain is 100 times longer than the estimated autocorrelation time. With this definition of convergence, chains converged after 120K and 270K steps, respectively, for the model 'GP-only' and 'Star-Planet' models. 

\begin{figure*}
\includegraphics[width=2\columnwidth]{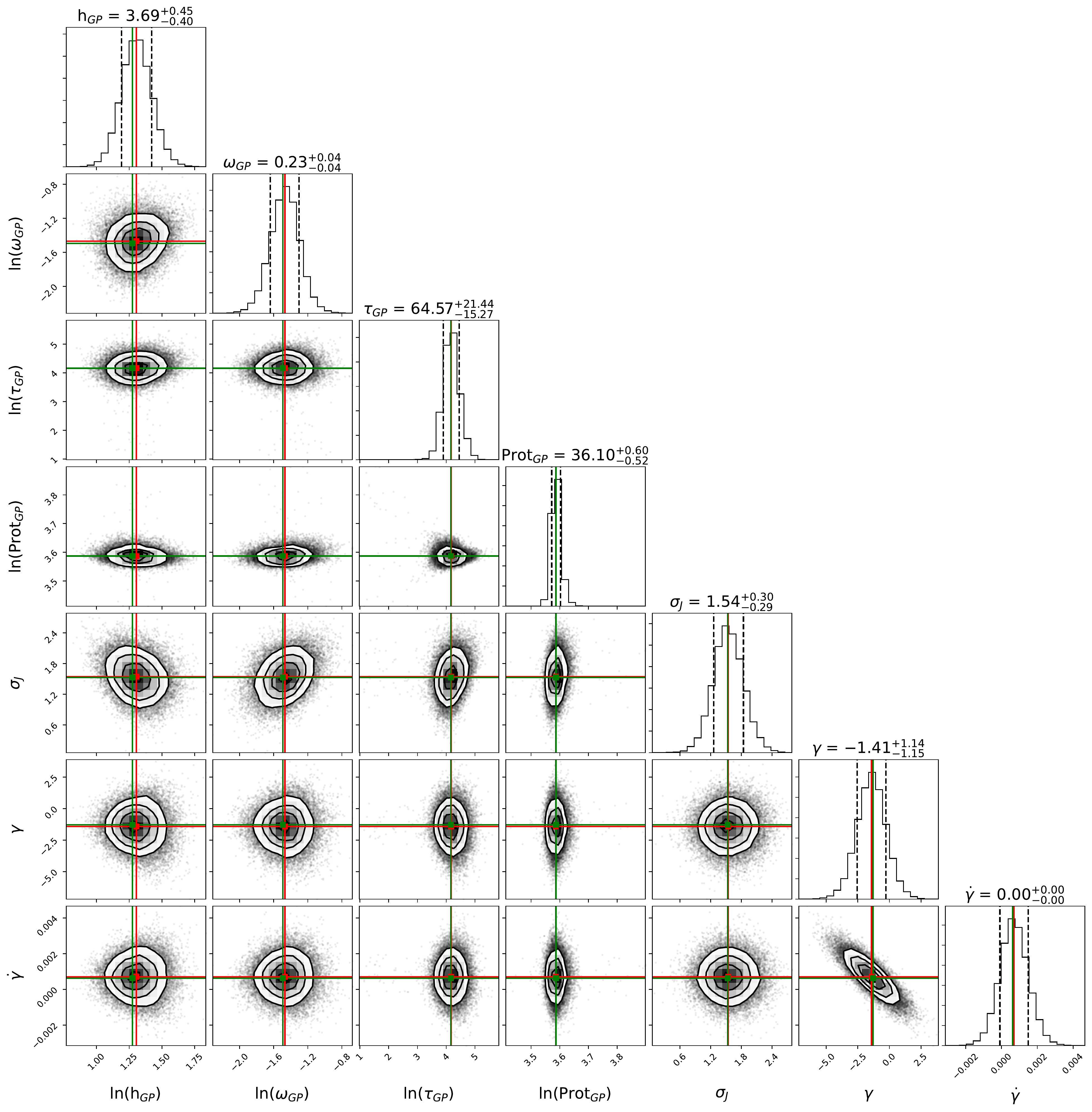}
\caption{Posterior distribution of the 'GP-only' model in which median and maximum a-posterior probability (MAP) have been marked (respectively, red and green line).}
\label{star_only}
\end{figure*}

\begin{figure*}
\includegraphics[width=2\columnwidth]{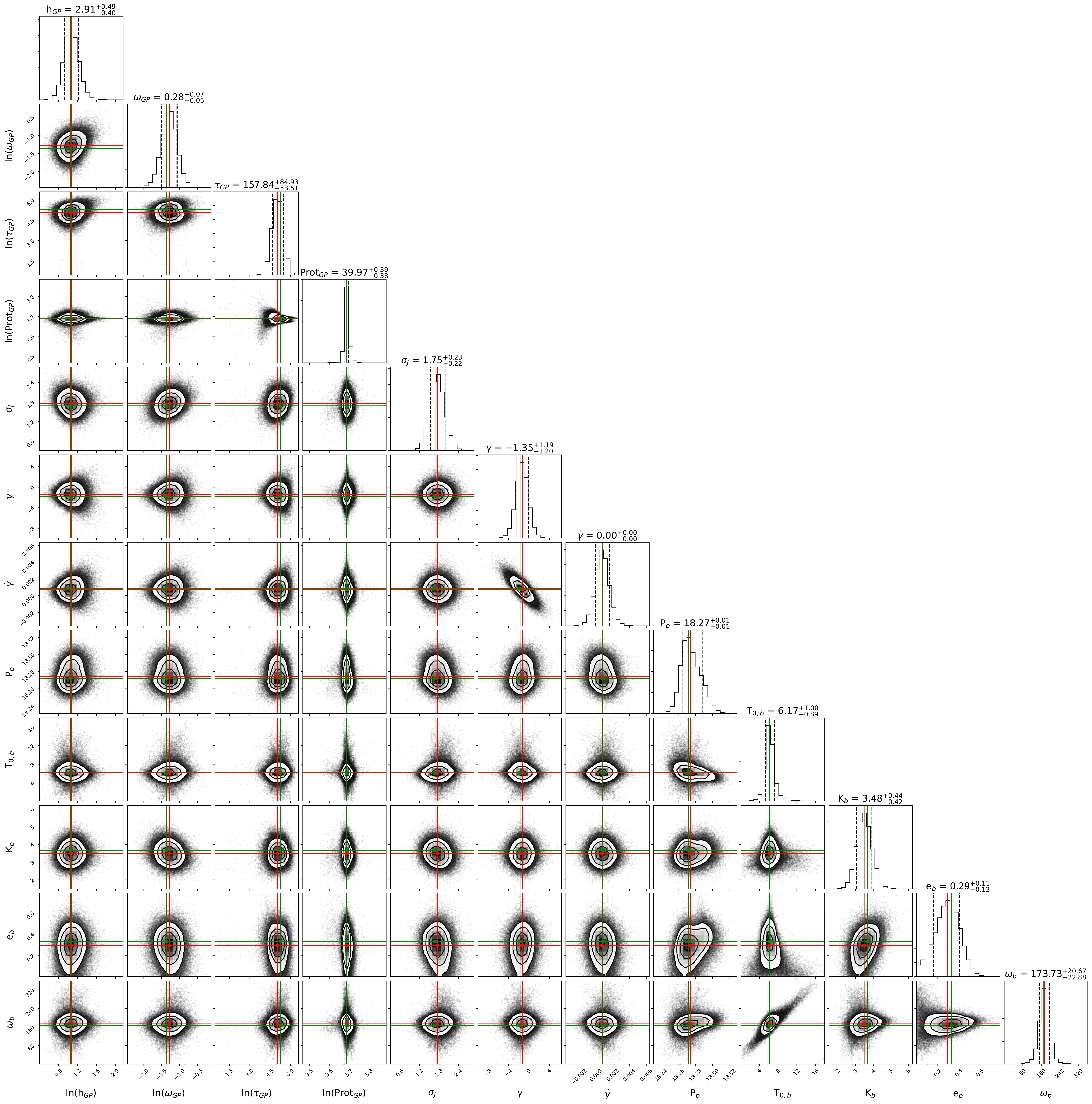}
\caption{Same as for Fig.\ref{star_only} but in the case of the 'Star-planet' model.}
\label{star_planet}
\end{figure*}

\begin{figure}[!htb]
\centering
\includegraphics[scale=0.35]{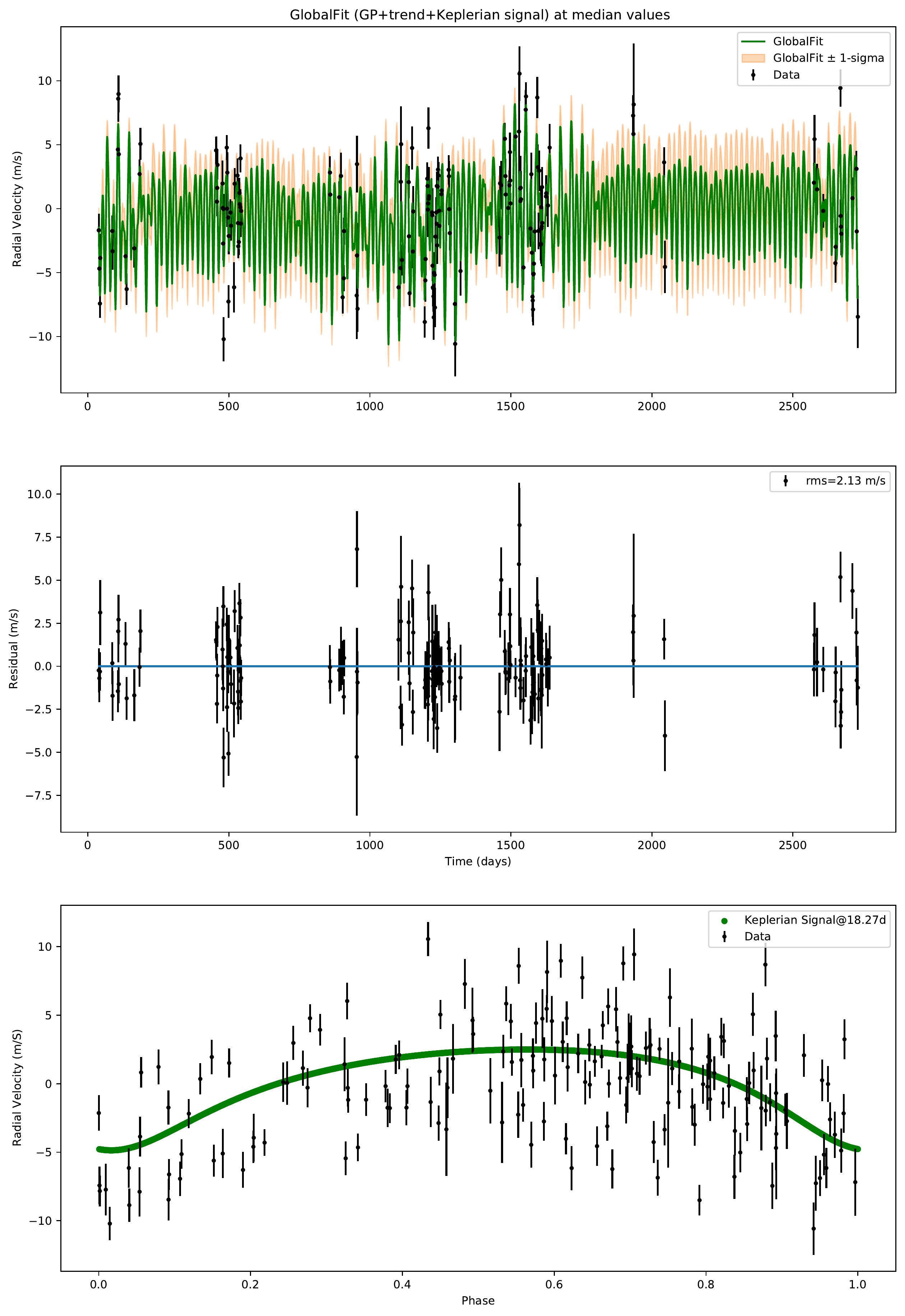}
\caption{
Best-fit 'Star-planet' model (top), RV residuals (middle), and RV curve folded at the
best-fit orbital period for GJ 9689 b (bottom).}
\label{best-fit-model}
\end{figure}

The resulting posterior distributions are presented, respectively, in Fig.~\ref{star_only} and Fig.~\ref{star_planet}. As an estimate of the goodness of the model, we have calculated the Bayesian Information Criterion (BIC), defined as follows
\begin{equation}
BIC = k \ln(N) - 2 \ln(\mathcal{L})
\end{equation}
where $k$ is the number of model parameters, $N$  the number of data points and $\mathcal{L}$  the maximum likelihood of the model. We obtain that there is strong evidence in supporting the 'Star-planet' model (BIC=912.4) against the 'GP-only' model (BIC=930.9) because the BIC difference is more than 10 \citep{bayes_factors}.
Figure~\ref{best-fit-model} shows the best-fit 'Star-planet' model (top), the corresponding RV residuals (middle), and the RV curve folded at the 
best-fit orbital period for the detected planet (bottom). The best-fit parameters are listed in Table~\ref{tab:priors}.
Given that the derived eccentricity is not statistically significant, we also provide the 68\% upper limit. 
We attempted a 'Star-planet' model with zero eccentricity. The corresponding BIC value is 
almost identical to the eccentric model.

 In order to test whether the GP part of the model can generate 
 a spurious signal at $\sim$ 18 d,
 we derive a FAP-like significance of the $\sim$ 18 d signal by drawing a 
 sample of 10$^{\rm 4}$ RV curves from the 'Star-planet' model using the best fit
 parameters but excluding the planetary part. Then, we counted how many samples have the power of the 
 $\sim$ 18.27 d signal greater than the value obtained by evaluating the periodogram on the original
 RV data. This condition was never achieved. 

\begin{table*}[t]
\centering
\caption{Best-fit values obtained for the 'Star-planet' model. 
Parameters $e$ and $\omega$ are derived from the explored parameters $\sqrt{e} \cos(\omega)$ and $\sqrt{e} \sin(\omega)$.}
\label{tab:priors}
\begin{tabular}{lllr}
\hline
\hline
\noalign{\smallskip}
Parameter & Prior & Description  & Best-fit value \\
\noalign{\smallskip}	
\hline	
\noalign{\smallskip}
\multicolumn{4}{l}{\textit{Linear trend}} \\
$\gamma$  (m s$^{\rm -1}$)     &  $\mathcal{U}$(-10, 10)           & Ordinate          & ${-1.35}_{-1.20}^{+1.19}$ \\
$\dot{\gamma}$ (m s$^{\rm -1}$)     &  $\mathcal{U}$(-10, 10)           & Slope             & ${0.00}_{0.00}^{0.00}$    \\
\noalign{\smallskip}							    
\multicolumn{4}{l}{\textit{GP parameters}} \\
\noalign{\smallskip}
$\sigma_{Jit}$    (m s$^{\rm -1}$) & $\mathcal{U}$($10^{-2}$, $10^2$)             & White noise term                        & ${1.75}_{-0.22}^{+0.23}$  \\
$h$              (m s$^{\rm -1}$) & $\mathcal{L} \mathcal{U}$($10^{-2}$, $10^2$) & Amplitude of the covariance             & ${2.91}_{-0.40}^{+0.49}$ \\
$\tau$           (d)             &  $\mathcal{L} \mathcal{U}$($1$, $10^5$)      & Timescale of the exponential component  & ${157.84}_{-53.51}^{+84.93}$ \\
$\omega$                         & $\mathcal{L} \mathcal{U}$($10^{-2}$, $10$)   & Weight of the periodic component        &              ${0.28}_{-0.05}^{+0.07}$\\
$P_{\rm rot}$     (d)             & $\mathcal{L} \mathcal{U}$(30, 50)            & Rotation period                         & ${39.97}_{-0.38}^{+0.39}$\\
\noalign{\smallskip}         
\multicolumn{4}{l}{\textit{Planet parameters} }\\	
\noalign{\smallskip}										    
$P_b$ (d)		    & $\mathcal{U}$ (15, 20) 	& Period                     & ${18.27}_{-0.01}^{+0.01}$   \\
$T_{0,b}$ (BJD-2,456,400 d) & $\mathcal{U}$ (0, 20)	& Time of periastron passage &  ${6.17}_{-0.89}^{+1.00}$   \\
$K_b$	(m s$^{\rm -1}$)    & $\mathcal{U}$ (0, 10 )  	& RV semi-amplitude          &  ${3.48}_{-0.42}^{+0.44}$  \\    				
$e_b$ 			    & $\mathcal{U}$ (0, 0.8 )	& Orbital eccentricity       & ${0.29}_{-0.13}^{+0.11}$, ($<$ 0.34)  \\
$\omega_b$ (deg) 	    & $\mathcal{U}$ (0, 360)	& Periastron angle           & ${173.73}_{-22.88}^{+20.67}$ \\
\noalign{\smallskip}	
\multicolumn{4}{l}{\textit{Derived quantities}}\\
\noalign{\smallskip}
$M_{\rm b}\sin i$ (M$_{\oplus}$) &  & Minimum mass            &  9.65 $\pm$ 1.41      \\
$a_{\rm b}$ (au)                 &  & Semi-major axis         &  0.1139 $\pm$ 0.0039  \\
$T_{\rm eq,b}$ (K)               &  & Equilibrium temperature &  413.88-492.19        \\
\hline	
\noalign{\smallskip}
\end{tabular}
\end{table*}

As a further investigation, we have also checked on the stability of the planetary signal by applying the same GP analysis with the 'Star-planet' model to the seasons defined in Sect.~\ref{subsec:seasons}. In this case we set the burn-in phase to 50K steps and we have run the second chain until it reaches 100K steps.

Due to the small amount of points in each season, we have shortened the prior ranges of the kernel and the linear trend parameters within 1$\sigma$ from the median of the posterior distributions presented in Fig.\ref{star_planet}. For the same reason the orbital period is constrained within 3$\sigma$ while the other prior ranges are unchanged with respect to the previous analysis. The result of this analysis is presented in Table \ref{tab:gp_to_seasons}.
Note that in all seasons we derive the same planetary parameters (within the uncertainties) which shows that
the Keplerian signal is coherent in all seasons.

\begin{figure}[htb]
\centering
\includegraphics[scale=0.5]{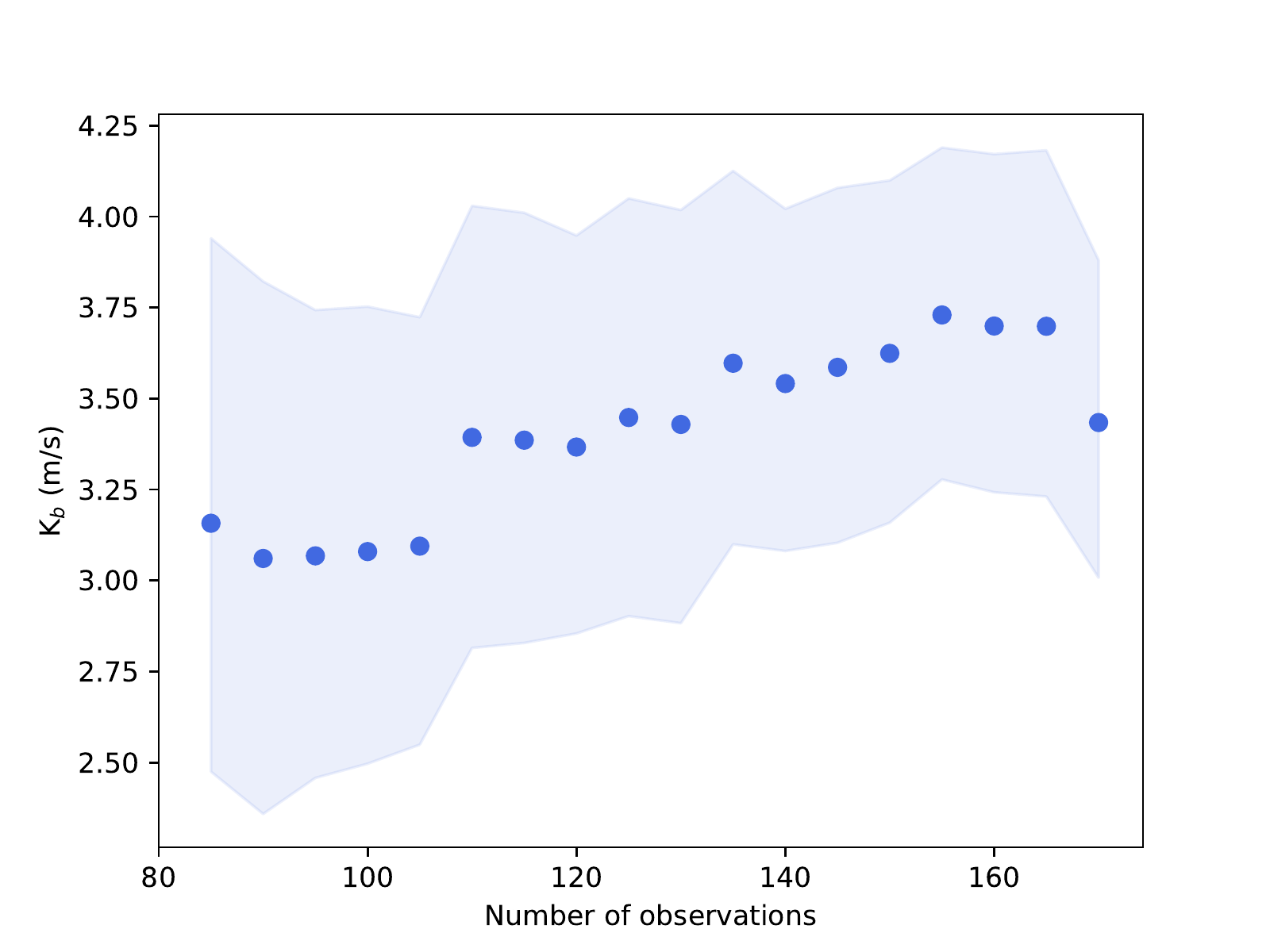}
\caption{RV semi-amplitude derived for GJ 9689 b using the 'Star-planet' model described in the text as a
function of the number of observations. The shadow region indicates the 1 $\sigma$ uncertainties.}
\label{kb_nobs}
\end{figure}

As an additional test on the stability of the planetary signal, we analyse how the RV semi-amplitude varies
as a function of the number of observations. 
The results are shown in Fig.~\ref{kb_nobs} where it can be seen that starting from $\sim$ 85 observations the value of $K_{\rm b}$ 
remains constant within the uncertainties around a value of 3.5 ms$^{\rm -1}$. Note that for this exercise we run
the 'Star-planet'  model with the priors as listed in Table~\ref{tab:priors} and run the second chain until it reaches 100K steps.

For the sake of completeness, we tested a model with a second additional planet.
Given that the RV time-series analysis reveals no more addition signals to the ones
already discussed, we performed a blind search, using a wide prior for the period of
the second planet. More specifically, we tested a model with a second planet with a period between 1 d and 100 d,
between 100 d and 200 d, and between 200 d and 300 d.
None of these models provide a lower BIC than the 'One planet - star' model, confirming that, if there are more planets in the system,
they are difficult to reveal with the data at hand.

 We  also run the 'Star-planet' model using the quasi-periodic with cosine (QPC) kernel
 defined in \cite{2021A&A...645A..58P}. The QPC kernel is defined as an QP kernel but it adds an additional term
  in order to account for the P$_{\rm rot}$/2 peaks in the autocorrelation function.
   The corresponding results are given in Table~\ref{tab:gp_to_seasons}. It can be seen that the
    best obtained values are almost identical to the ones derived using the QP.
     The QPC has a BIC value of 917.6, which is slightly larger than the value obtained using the QP kernel (BIC = 912.4).

\begin{table*}
\centering
\caption{
Best-fit values obtained for the 'Star-planet' model when using the full dataset of RVs, and the data corresponding
to the different seasons analysed in this work. The best-fit values obtained using the full RVs dataset
and the QPC kernel are also given. 
Parameters $e$ and $\omega$ are derived from the posterior distribution of $\sqrt{e} \cos(\omega)$ and $\sqrt{e} \sin(\omega)$. }
\label{tab:gp_to_seasons}
\begin{tabular}{l c c c c c}

\hline
\hline
\noalign{\smallskip}
Parameter & Full dataset & Season 1 & Season 2 & Season 3  & Full dataset (QPC) kernel \\
\noalign{\smallskip}
\hline
$P_b$ (d)        &${18.27}_{-0.01}^{+0.01}$    & --                         & --                       & --                       & ${18.27}_{-0.01}^{+0.01}$ \\
$T_{0,b}$ (d)     & ${6.17}_{-0.89}^{+1.00}$    & ${6.25}_{-2.03}^{+5.63}$    & ${7.61}_{-1.83}^{+2.00}$    & ${7.69}_{-2.59}^{+2.80}$   & ${6.19}_{-0.90}^{+1.05}$\\
$K_b$ (m/s)      & ${3.48}_{-0.42}^{+0.44}$    & ${3.32}_{-0.75}^{+0.77}$     & ${3.53}_{-0.74}^{+0.88}$    & ${3.71}_{-0.92}^{+0.98}$   & ${3.49}_{-0.43}^{+0.44}$\\
$e_b$            & ${0.29}_{-0.13}^{+0.11}$     & ${0.33}_{-0.21}^{+0.23}$    & ${0.32}_{-0.20}^{+0.20}$     & ${0.33}_{-0.17}^{+0.15}$   & ${0.29}_{-0.13}^{+0.11}$\\
$\omega_b$ (deg) & ${173.73}_{-22.88}^{+20.67}$ & ${167.67}_{-72.98}^{+48.90}$ & ${178.17}_{-31.74}^{+37.16}$ & ${177.19}_{-43.97}^{+49.47}$ &${173.47}_{-23.82}^{+20.80}$\\
\hline	
\noalign{\smallskip}
\end{tabular}
\end{table*}	

 As a final test, we also explored the full (hyper)-parameter space
 using the publicly available Monte Carlo (MC) nested sampler and Bayesian inference tool
 \textsc{MultiNest v3.10}  \citep[e.g.][]{Feroz2019}, through the \textsc{pyMultiNest} wrapper \citep{Buchner2014}.
 \textsc{MultiNest} is known to provide accurate estimates of the Bayesian evidence $\mathcal{Z}$,
 that can be used to perform a statistical comparison between different models.
 To test the planetary nature of the  $\sim$18 day signal (circular orbital approximation), we
 fitted the RVs using two different GP models. For the first model, we adopted only 
 a QPC kernel, while the second model is represented by the combination
 of a QP kernel and a sinusoid. For the hyper-parameter representing the
 stellar rotation period, which is the same both in the QPC and QP GP kernels, we used the uninformative
 prior $\mathcal{U}$(20,50) days. Both models are assumed to have an equal a-priori likelihood. As in our
 previous analyses we
 found that the QP plus planetary model is strongly favoured over the 'GP-only' QPC model
 ($\Delta\ln\mathcal{Z}$=+7.3, corresponding to an odds ratio of $\sim$1500:1), following
 the convention usually adopted for model selection \citep[e.g.][Table~1]{feroz2011}.
 This result shows that the $\sim$18 day signal is much better fitted by a sinusoid rather
 than being modelled through a QPC kernel as the first harmonic of the stellar rotation period. 

\section{Discussion}\label{others}
%
 
\begin{figure}[htb]
\centering
\includegraphics[scale=0.5]{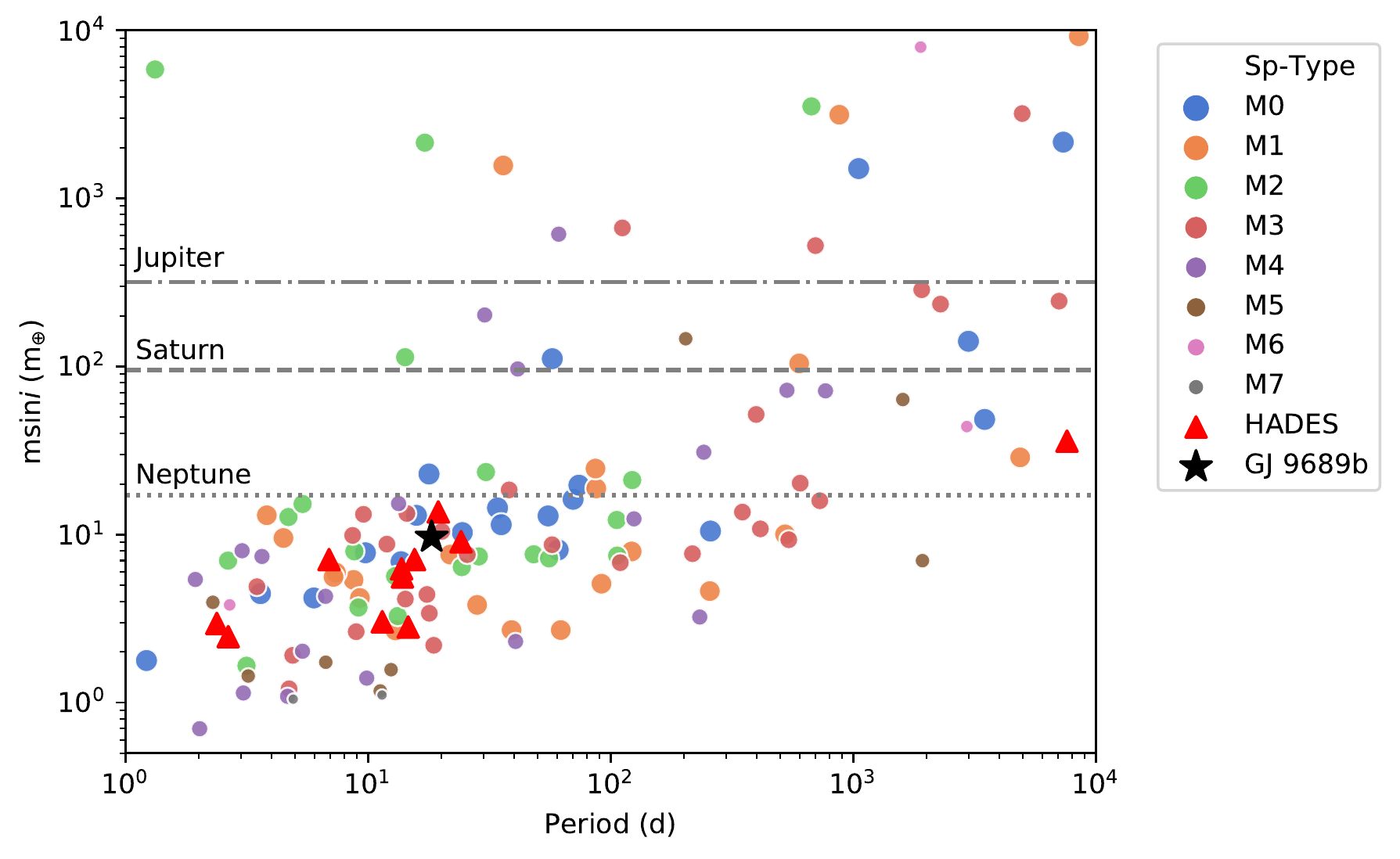}
\caption{Known radial velocity planets (planetary mass vs. orbital period diagram) around M dwarfs
(As listed at http://exoplanet.eu/ in December 2020).
Planets discovered by the HADES survey are shown as red triangles.
The planet GJ 9689 b is shown as a black star.}
\label{dist_acum_low_mass_lc}
\end{figure}

 From the best values of $K_{\rm b}$ and $P_{\rm b}$ derived in the previous section
 we derive a minimum mass for GJ 9689 b of M$_{\rm P}\sin i$ = 9.65 $\pm$ 1.41 M$_{\oplus}$
 and a semi-major axis a = 0.1139 $\pm$ 0.0039 au \citep[for the formulae see e.g.][Eqn. 1, 3]{1999ApJ...526..890C}.
 GJ 9689 b is therefore a super-Earth or a mini-Neptune like planet.
 Figure~\ref{dist_acum_low_mass_lc} shows the position of GJ 9689 b in the
 planetary mass vs. period diagram. For comparison purposes, the location of the known
 (radial velocity) planets around M dwarf stars are shown. 
 It can be seen that GJ 9689 b has a period and a minimum mass similar to other
 HADES planets. 
 Note that most of the HADES discoveries have periods shorter than 20 d 
 and minimum masses lower than the mass of Neptune.
 Only one HADES planet, namely GJ 15 Ac, has a higher mass and an orbit at a wide distance from its host star,
 P $\sim$ 7600 d \citep{2018A&A...617A.104P}. 
 This is in agreement with other
 radial velocity surveys that found that the frequency of gas-giant planets
 around M stars is lower than that around solar-type hosts
 \citep{2003AJ....126.3099E,2006ApJ...649..436E,2006PASP..118.1685B,2007A&A...474..293B,2008PASP..120..531C,2010PASP..122..149J}.

 Following the definition of Habitable Zone (HZ) of \cite{2013ApJ...765..131K},
 the inner edge of the HZ for GJ 9689 was computed with the most optimistic limits
 (recent Venus) which corresponds to a semi-major axis of a$_{\rm HZ}$ = 0.20 au,
 which is larger than the orbit of GJ 9689 b. 
 The equilibrium temperature of the planet can be determined from the balance
 between the incident radiation from the host star, and that absorbed by the planet
 (or by its atmosphere). It can be written as:

\begin{equation}
T_{eq}=T_{\star}\left(\frac{R_{\star}}{2a}\right)^{1/2}\left[f\left(1-A_{B}\right)\right]^{1/4} 
\end{equation}
\noindent where additional heat sources (such as the greenhouse effect) are not taken into account. 
In this equation $A_{B}$ is the Bond albedo and $f$ the heat redistribution factor.
The value of $f$ goes from $f$ = 1 for an isotropic planetary emission, to $f$ = 2 
when only the day-side re-radiates the energy absorbed, as it could be the case for tidally-locked
planets without oceans or atmosphere \citep{2005ApJ...626..523C,2017ApJ...837L...1M}.
An upper limit on $T_{eq}$ may be obtained by setting $A_{B}$ = 0. 
For GJ 9689 b we find upper limits on $T_{eq}$ between 413.88 K ($f$ = 1) and 492.19 K ($f$ = 2).
 Therefore, GJ 9689 b have a $T_{eq}$ which is $\sim$ 140-240 K higher than the equilibrium temperature of
 the rocky planets in our Solar System. For example Venus has $T_{eq}$ $\sim$ 230 K, while
 the Earth has a value of $\sim$ 255 K, and for Mars we have $T_{eq}$ $\sim$ 212 K \citep{2018exha.book.....P}.
 However, the equilibrium temperature of GJ 9689 b is
 in agreement with the values derived for other super-Earth
 planets found around M dwarfs like GJ 3998 c, $\sim$ 420 K, or  Gl 886 b, $\sim$ 379-450 K
 \citep[][]{2016A&A...593A.117A,2019A&A...622A.193A}.

 The composition of exoplanets is an important but highly problematic issue.
 To start with, error uncertainties in stellar radius and mass are usually
 large and lower values of mass and radius for the host star translate to overall higher densities 
 \citep[e.g.][]{2020A&A...641A.113M}.
 Even if the planet transits, and accurate planetary mass and radius measurements are available,
 the composition of low-mass planets is plagued with degeneracies
 that arise mainly from trade-offs between the different composition building blocks
 \citep[e.g.][]{2007ApJ...656..545V,2010ApJ...712..974R,2020MNRAS.499..932P}.
 
\begin{figure}[htb]
\centering
\includegraphics[scale=0.62]{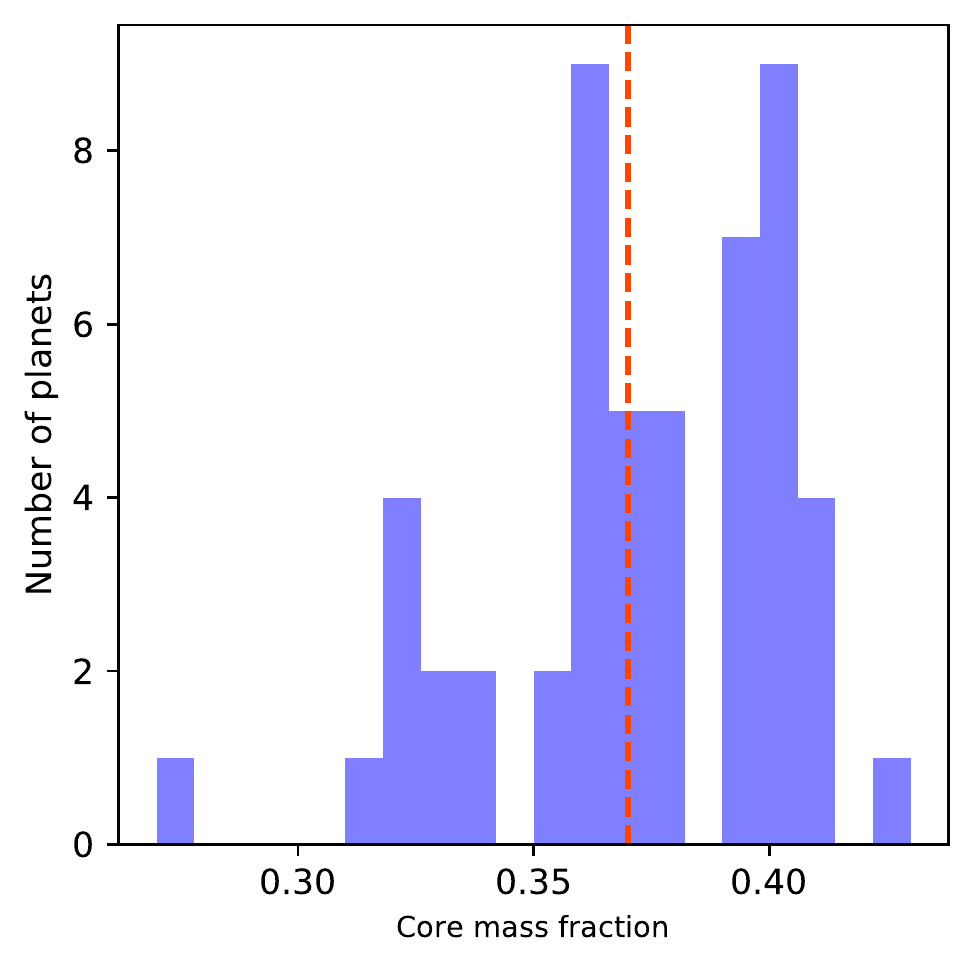}
\caption{Histogram of derived CMF for suspected rocky planets around M dwarfs.
The vertical red line shows the median of the distribution.}
\label{cmf_histogram}
\end{figure}

 In order to solve this problem, several works have suggested to use
 the abundances of refractory elements 
 of the host star
to constrain the refractory content of the planet, which
 is likely a good approximation from a statistical approach \citep[e.g.][]{2015A&A...577A..83D,2017ApJ...850...93B,2017A&A...608A..94S}.
 Recently, \cite{2020A&A...644A..68M} determine the stellar abundances of a large sample of M dwarfs for several elements different from iron
 analysing high-resolution optical spectra.
 Using these abundances, we estimated the core mass fraction (CMF) of 52  
 rocky exoplanets (planetary masses between 1 and 20 M$_{\oplus}$) around M dwarfs.
 We follow the CMF definition as provided in \cite{2020arXiv201108893S}
 that assumes that the planetary core is pure iron and the mantle reflects fully oxidised Mg and Si.
 The results are shown in Fig.~\ref{cmf_histogram} where the histogram of the derived CMFs
 is shown. 
 We find that the CMF values of small planets around M dwarfs varies from 0.27 to 0.52 with a median
 value of 0.38. According to our results, GJ 9689 b have a CMF value 0.34, close to the median
 of the distribution.
 It is worth noticing that these values are slightly larger than the CMF values derived for 11 transiting planets around
 FGK stars (median CMF = 0.29) in \cite{2020arXiv201108893S}. This trend, if confirmed, may indicate that small planets around M dwarfs
 might have larger cores. 

  Finally, we should note that the derived timescale of active regions is shorter in the 'GP-only' model than in the
  'Star-planet'.  While the former, $\tau_{\rm GP}$ = 64.57$^{\rm + 21.44}_{\rm - 15.27}$ d, is consistent with roughly two rotation periods,
  the latter, $\tau_{\rm GP}$ = 157.84$^{\rm + 84.93}_{\rm - 53.51}$ d, is much larger (around 4.0 rotation periods) and consistent with the typical
  active regions lifetimes derived for other M dwarf stars \citep{2017A&A...598A..28S,2019A&A...625A.126P}.
 To further investigate this issue, we run our 'GP-only' model for the S-index time series and
 the V band EXORAP photometry  although the results do not allow us to reach a clear conclusion.
 Results for the S-index  
 provide a typical
 active regions timescale of $\tau_{\rm GP}$ = 63.87$^{\rm + 79.64}_{\rm - 24.38}$ d,
 although the distribution is rather broad and has a tail towards larger values. 
 However, the analysis of the V band photometry
 returns a correlation decay timescale of $\tau_{\rm GP}$ = 38.27$^{\rm + 41.91}_{\rm - 8.07}$ d  which is comparable to the stellar rotation period, but a secondary peak appears at $\tau_{\rm GP}$ $\sim$ 100 d.
 The corresponding corner plots can be seen in Fig.~\ref{star_only_activity} (S-index fit) and in Fig.~\ref{star_only_photometry} (EXORAP photometry). 

  GJ 9689 is expected to be observed by the {\it Tess} mission \citep{2015JATIS...1a4003R} in 2022, Sector 54
  (July 9 to August 5, in cycle 4). In order to get an estimate of the transit probability and depth for GJ 9689 b,
  a value of the planetary radius is needed. 
  Using the probabilistic mass–radius relationship by \cite{2017ApJ...834...17C}, we obtain for GJ 9689 b a radius of
  R$_{\rm P}$ = 2.92$^{\rm + 1.05}_{\rm - 0.58}$ R$_{\oplus}$ which implies a geometric transit probability of
  2.5\% and a transit depth of 0.23\%. 
  Although the transit probability is rather low, a potential transit might provide a constraining point in the mass-radius diagram of known planets and also enable determining the bulk density. A comparison of the CMF derived 
   by assuming that the planet reflects the host star's major rocky building elemental abundances
   and the CMF value derived from the planetary bulk density
   might help us to unravel how GJ 9689 b formed and evolved. 

  Astrometric measurements are another source of information that might be used for the
  characterisation of the GJ 9689 planetary architecture. 
  {\it Gaia} EDR3 data \citep{2020yCat.1350....0G} shows that 
  GJ 9689 has an astrometric noise of 73$\mu$as (with a statistical significance of 6.5 $\sigma$) 	
  and a Renormalised Unit Weight Error (RUWE) value of 0.996. 
  Furthermore, no significant proper motion anomaly between {\it Gaia} DR2 and {\it Hipparcos}
  has been reported \citep{2019A&A...623A..72K}.
  Therefore, it is plausible that there are no massive companions at long periods.
  Detailed RV detection limits for the stars observed within the framework of the HADES survey will be
  addressed in a forthcoming work.

  It is worth noticing that GJ 9689 b is not the first planet
  around an M dwarf with an orbital period close to half
  the stellar rotation period. Some examples include Gl 49 b \citep{2019A&A...624A.123P} or
  GJ 720 Ab \citep{2021arXiv210309643G}. 
  The detection of such planets is challenging. 
  It is known that dark spots in the stellar surface might produce a RV modulation with a period of P$_{\rm rot}$/2,
  while the induced RV period due to the suppression of the surface convection is equal to P$_{\rm rot}$
  \citep[see e.g.][]{2010A&A...520A..53L}.
  The period found in the RV might be slightly different from P$_{\rm rot}$/2 due to several phenomena such as 
  differential rotation or the evolution of the spot in the active longitude.
  While lifetimes of active longitudes are not well known, the possibility an of active longitude with time scales
  of several years might not be excluded in M dwarfs. Note that single spots may survive even for hundreds of days
  in the surface of M dwarfs \citep[e.g.][]{2017MNRAS.472.1618G}.
  In the case of GJ 9689, given its rotational period and the photometric variability of the star,
  and assuming a stable active longitude covered by
  dark spots with an area of 1\% of the stellar disc, we derive a semi-amplitude of $\sim$ 7.2 ms$^{\rm -1}$ for the 
  signal due to the magnetic activity of such spots (which is larger than the observed RV variability). 
 
  This consideration and all the analysis performed in this work make us confident that the signal at 18.27 d found
  in the RV time-series of GJ 9689 is truly Keplerian in nature. 
  An 'ultimate' confirmation of planets like GJ 9689 b will likely be possible in a near future 
  if high-resolution spectrographs in the near infrared domain are able to reach enough RV precision.
  

\section{Conclusions}\label{conclusions} 

 Understanding the origin and evolution of stars and planetary systems
 is one of the major goals in modern Astrophysics. M dwarfs have emerged
 as promising targets in the search for small planets. Being less massive
 and cooler than solar-type stars,  planets 
 have a larger radial velocity amplitude and
 their habitable zones are closer to the host star \citep[e.g.][]{2010ApJ...710..432R}.
 However, the expected periods of 
 planets in the habitable zone
 around M stars 
 may be comparable
 to the rotation period of the host star. As a consequence, unravelling
 the origin (stellar or truly Keplerian) of periodic signals in the RV dataset of M dwarfs 
 is a complicated task.  
 
 The nearby M dwarf GJ 9689 is an example of this problematic. 
 In this work we have analysed more than seven years of RV data points of GJ 9689.
 The radial velocity analysis reveals two signals at 39.31 d and 18.21 d. 
 A detailed study of the main activity indexes and photometry 
 allows us to identify the 39.31 d as the rotation period of the star.
 This hypothesis is confirmed by studying the stability of this signal as a function of the
 number of observations, epoch of observations, and spectral range  used to compute
 the radial velocity.

 A careful analysis of the spectral window in several observing seasons shows that the
 18.27 d signal is not an harmonic or an alias of the 39.32 d period. In addition, activity indexes,
 CCF asymmetry indicators, and photometric data show no 
 variability at all at 18.27 d. The signal is stable in all analysed epochs, it does not
 show variations with the number of observations, and it is also visible even if only
 the reddest part of the spectra are used in the computation of the radial velocity dataset.
 We therefore conclude that the 18.27 d signal is most likely truly Keplerian in nature.
 In order to derive the minimum mass and orbital parameters of
 GJ 9689 b we fitted the RV time series with a Keplerian model
 combined with a GP quasi-periodic model to take into account
 the stellar activity signal. We obtain a period P = 18.27 $\pm$ 0.01 d,
 a semi-major axis of a = 0.114 $\pm$ 0.004 au, and a minimum mass of
 M$_{\rm P}\sin i$ = 9.65 $\pm$ 1.41 M$_{\oplus}$.
 Assuming that the composition of a rocky planet directly mirrors the relative Fe, Mg, and Si abundances of its host
 we derived the CMF of several suspected rocky planets around M dwarfs, finding that the CMF value of GJ 9689 b is close to the median
 value of the distribution. 

 GJ 9689 b joins the population of Super-Earths at short periods (P $<$ 20 d)
 around M dwarfs that is currently emerging (see Fig.~\ref{dist_acum_low_mass_lc}). Further studies of the properties of these
 planets 
 will help us to explore
 a regime of protoplanetary disc and host stars conditions very different from FGK stars,
 providing additional constraints for planet formation models.

\begin{acknowledgements}

J.M., and G.M. acknowledge support from the 
accordo attuativo ASI-INAF n.2021-5-HH.0 ``Partecipazione italiana alla fase B2/C della missione Ariel''.
A.P., A.M., M.P., and A.S. acknowledge partial contribution from the agreement ASI-INAF n.2018-16-HH.0.
E.G.A. acknowledge support from the Spanish Ministery for Science, Innovation, and Universities through projects AYA-2016-79425-C3-1/2/3-P, AYA2015-69350-C3-2-P, ESP2017-87676-C5-2- R, ESP2017-87143-R. The Centro de Astrobiolog\'ia (CAB, CSIC-INTA) is a Center of Excellence ``Maria de Maeztu''. 
M.P and I.R acknowledge support from the Spanish Ministry of Science and Innovation and the European Regional Development Fund through grant PGC2018-098153-B- C33, as well as the support of the Generalitat de Catalunya/CERCA programme.
J.I.G.H. acknowledges financial support from Spanish Ministry of Science and Innovation (MICINN) under the 2013 Ram\'on y Cajal program RYC-2013- 14875. B.T.P. acknowledges Fundaci\'on La Caixa for the financial support received in the form of a Ph.D. contract. A.S.M. acknowledges financial support from the  Spanish MICINN under the 2019 Juan de la Cierva Programme.  B.T.P., A.S.M., J.I.G.H., R.R. acknowledge financial support from the Spanish MICINN AYA2017-86389-P.
This work has made use of data from the European Space Agency (ESA)
mission {\it Gaia} (\url{https://www.cosmos.esa.int/gaia}), processed by
the {\it Gaia} Data Processing and Analysis Consortium (DPAC,
\url{https://www.cosmos.esa.int/web/gaia/dpac/consortium}). Funding
for the DPAC has been provided by national institutions, in particular
the institutions participating in the {\it Gaia} Multilateral Agreement.

\end{acknowledgements}

%
\bibliographystyle{aa}
\bibliography{M059_repor.bib}


\begin{appendix}

\section{Additional tables}

Table~\ref{M059_log} provides the observational data collected with the HARPS-N spectrograph for GJ 9689 and used in the present study.
We list the observations dates (in barycentric Julian date or BJD), the RVs, and activity S, H$\alpha$, 
and Na~{\sc i} D$_{\rm 1}$, D$_{\rm 2}$ indexes with their corresponding uncertainties.

\longtab[1]{
\begin{longtable}{ccccccccc}
\caption{
 { Observation log of GJ 9689}
  }\label{M059_log}\\
\hline
BJD - 2400000   &   RV                &  $\Delta$RV      &  S-index &  $\Delta$(S-index) & H$\alpha$-index & $\Delta$(H$\alpha$-index) & Na~{\sc i}-index & $\Delta$(Na~{\sc i}-index)\\ 
(d)             &  (ms$^{\rm -1}$)    &  (ms$^{\rm -1}$) &          &                  &           & & &       \\
\hline
\endfirsthead
\caption{Continued.} \\
\hline
BJD - 2400000   &   RV                &  $\Delta$RV     &  S-index &  $\Delta$(S-index) & H$\alpha$-index & $\Delta$(H$\alpha$-index) & Na~{\sc i}-index & $\Delta$(Na~{\sc i}-index)\\ 
(d)            &  (ms$^{\rm -1}$)    &  (ms$^{\rm -1}$) &          &                  &     & & &             \\
\hline
\endhead
\hline
\endfoot
\hline
\endlastfoot
56438.684901	&	-1.69	&	1.29	&	1.1905	&	0.0098	&	0.3444	&	0.0008	&	0.1274	&	0.0008	\\
56440.698234	&	-4.69	&	1.39	&	1.0702	&	0.0127	&	0.3420	&	0.0009	&	0.1212	&	0.0010	\\
56442.687720	&	-7.42	&	1.12	&	1.0101	&	0.0093	&	0.3381	&	0.0008	&	0.1219	&	0.0008	\\
56443.663855	&	-3.86	&	1.88	&	0.9711	&	0.0211	&	0.3369	&	0.0013	&	0.1234	&	0.0015	\\
56486.620666	&	-1.75	&	1.22	&	1.1106	&	0.0133	&	0.3395	&	0.0010	&	0.1219	&	0.0010	\\
56487.580888	&	-3.33	&	1.45	&	1.1052	&	0.0142	&	0.3388	&	0.0010	&	0.1198	&	0.0010	\\
56506.483363	&	4.62	&	1.22	&	1.1593	&	0.0115	&	0.3400	&	0.0009	&	0.1250	&	0.0009	\\
56507.593975	&	8.59	&	1.79	&	1.3207	&	0.0182	&	0.3484	&	0.0009	&	0.1226	&	0.0011	\\
56508.608809	&	8.97	&	1.45	&	1.2414	&	0.0127	&	0.3397	&	0.0008	&	0.1237	&	0.0009	\\
56509.619800	&	4.26	&	1.11	&	1.3287	&	0.0138	&	0.3498	&	0.0009	&	0.1274	&	0.0010	\\
56533.476873	&	-3.72	&	1.27	&	0.9790	&	0.0099	&	0.3389	&	0.0008	&	0.1201	&	0.0008	\\
56537.502695	&	-6.29	&	1.24	&	1.0663	&	0.0093	&	0.3391	&	0.0007	&	0.1179	&	0.0007	\\
56564.543673	&	-3.11	&	1.51	&	1.2143	&	0.0157	&	0.3414	&	0.0010	&	0.1237	&	0.0011	\\
56583.391762	&	2.71	&	1.13	&	1.0735	&	0.0104	&	0.3399	&	0.0008	&	0.1220	&	0.0008	\\
56586.321128	&	5.07	&	1.25	&	1.2484	&	0.0128	&	0.3459	&	0.0007	&	0.1230	&	0.0009	\\
56854.577773	&	4.55	&	1.08	&	1.2765	&	0.0102	&	0.3419	&	0.0007	&	0.1283	&	0.0008	\\
56855.567308	&	4.57	&	1.07	&	1.2634	&	0.0104	&	0.3467	&	0.0008	&	0.1292	&	0.0008	\\
56857.677235	&	0.55	&	1.14	&	1.2984	&	0.0116	&	0.3482	&	0.0007	&	0.1262	&	0.0008	\\
56858.630655	&	1.62	&	1.26	&	1.2757	&	0.0130	&	0.3448	&	0.0008	&	0.1266	&	0.0009	\\
56859.643762	&	3.42	&	1.16	&	1.2015	&	0.0147	&	0.3453	&	0.0008	&	0.1239	&	0.0009	\\
56877.602491	&	1.97	&	1.79	&	1.2235	&	0.0243	&	0.3424	&	0.0012	&	0.1282	&	0.0015	\\
56878.585884	&	0.05	&	1.07	&	1.1074	&	0.0097	&	0.3417	&	0.0008	&	0.1246	&	0.0008	\\
56879.494416	&	-2.73	&	1.31	&	1.1120	&	0.0119	&	0.3426	&	0.0010	&	0.1236	&	0.0010	\\
56880.485352	&	-0.03	&	1.18	&	1.0715	&	0.0109	&	0.3467	&	0.0009	&	0.1260	&	0.0009	\\
56881.470848	&	-10.21	&	1.73	&	1.2427	&	0.0200	&	0.3501	&	0.0013	&	0.1282	&	0.0013	\\
56892.466346	&	4.78	&	0.97	&	1.2800	&	0.0093	&	0.3464	&	0.0007	&	0.1257	&	0.0007	\\
56893.481535	&	0.00	&	1.43	&	1.3666	&	0.0176	&	0.3471	&	0.0011	&	0.1264	&	0.0012	\\
56894.484073	&	2.82	&	1.58	&	1.3408	&	0.0151	&	0.3501	&	0.0010	&	0.1268	&	0.0011	\\
56897.506643	&	-0.69	&	1.24	&	1.3410	&	0.0157	&	0.3478	&	0.0012	&	0.1274	&	0.0011	\\
56898.464617	&	-7.26	&	1.28	&	1.2680	&	0.0138	&	0.3480	&	0.0010	&	0.1266	&	0.0010	\\
56899.477380	&	-2.14	&	1.44	&	1.3657	&	0.0173	&	0.3507	&	0.0012	&	0.1250	&	0.0012	\\
56905.470066	&	-0.30	&	1.02	&	1.2288	&	0.0109	&	0.3431	&	0.0008	&	0.1259	&	0.0008	\\
56907.464509	&	-1.34	&	1.15	&	1.1404	&	0.0104	&	0.3410	&	0.0008	&	0.1233	&	0.0008	\\
56918.472879	&	-6.15	&	1.97	&	1.1111	&	0.0237	&	0.3420	&	0.0014	&	0.1319	&	0.0016	\\
56920.469597	&	1.95	&	1.23	&	1.1492	&	0.0128	&	0.3441	&	0.0009	&	0.1233	&	0.0010	\\
56928.458456	&	14.93	&	3.72	&	1.0998	&	0.0441	&	0.3423	&	0.0012	&	0.1202	&	0.0021	\\
56930.505522	&	2.59	&	1.34	&	1.1514	&	0.0126	&	0.3390	&	0.0008	&	0.1236	&	0.0009	\\
56932.461632	&	-1.13	&	1.38	&	1.1538	&	0.0189	&	0.3406	&	0.0012	&	0.1239	&	0.0013	\\
56933.355305	&	-2.94	&	0.93	&	1.2501	&	0.0117	&	0.3435	&	0.0008	&	0.1256	&	0.0008	\\
56935.349440	&	-2.61	&	1.01	&	1.3033	&	0.0097	&	0.3440	&	0.0007	&	0.1270	&	0.0007	\\
56937.459316	&	1.23	&	1.19	&	1.2586	&	0.0113	&	0.3444	&	0.0008	&	0.1282	&	0.0008	\\
56938.464540	&	0.35	&	1.18	&	1.2257	&	0.0108	&	0.3477	&	0.0007	&	0.1293	&	0.0008	\\
56940.460277	&	0.11	&	1.40	&	1.3123	&	0.0120	&	0.3471	&	0.0009	&	0.1292	&	0.0009	\\
56941.347152	&	3.93	&	1.08	&	1.1803	&	0.0087	&	0.3448	&	0.0007	&	0.1277	&	0.0007	\\
56942.454785	&	-1.17	&	1.06	&	1.2207	&	0.0110	&	0.3438	&	0.0007	&	0.1246	&	0.0008	\\
56943.451164	&	-0.18	&	1.06	&	1.2204	&	0.0109	&	0.3462	&	0.0007	&	0.1243	&	0.0008	\\
57258.456713	&	2.82	&	1.28	&	1.2957	&	0.0127	&	0.3509	&	0.0008	&	0.1283	&	0.0009	\\
57259.479424	&	1.10	&	1.30	&	1.5259	&	0.0171	&	0.3656	&	0.0010	&	0.1341	&	0.0011	\\
57291.395613	&	0.90	&	1.24	&	1.3580	&	0.0123	&	0.3555	&	0.0009	&	0.1310	&	0.0009	\\
57297.467722	&	2.55	&	1.84	&	1.2680	&	0.0220	&	0.3441	&	0.0010	&	0.1198	&	0.0013	\\
57303.423054	&	-6.93	&	1.28	&	1.1434	&	0.0112	&	0.3463	&	0.0008	&	0.1276	&	0.0009	\\
57307.409630	&	-5.44	&	1.02	&	1.0423	&	0.0098	&	0.3371	&	0.0009	&	0.1230	&	0.0008	\\
57308.417641	&	-1.76	&	1.06	&	1.1055	&	0.0093	&	0.3370	&	0.0008	&	0.1225	&	0.0008	\\
57352.336645	&		&		&	0.5182	&	0.2654	&	0.3394	&	0.0109	&	0.1491	&	0.0310	\\
57353.310991	&	-6.79	&	3.41	&	1.3845	&	0.0587	&	0.3437	&	0.0022	&	0.1221	&	0.0032	\\
57354.304734	&	3.48	&	2.22	&	1.2338	&	0.0448	&	0.3410	&	0.0019	&	0.1252	&	0.0025	\\
57354.317418	&	-3.66	&	2.54	&	1.0779	&	0.0421	&	0.3435	&	0.0020	&	0.1287	&	0.0026	\\
57356.308205	&	-7.83	&	1.82	&	1.1784	&	0.0203	&	0.3392	&	0.0012	&	0.1230	&	0.0013	\\
57501.700690	&	-6.15	&	2.37	&	1.3670	&	0.0415	&	0.3617	&	0.0018	&	0.1388	&	0.0024	\\
57508.699547	&	-4.65	&	1.26	&	1.3434	&	0.0178	&	0.3489	&	0.0008	&	0.1259	&	0.0011	\\
57509.691609	&	2.10	&	2.39	&	1.1425	&	0.0303	&	0.3446	&	0.0012	&	0.1248	&	0.0017	\\
57510.683844	&	5.04	&	2.95	&	1.2858	&	0.0266	&	0.3399	&	0.0010	&	0.1210	&	0.0014	\\
57513.706098	&	-4.02	&	1.22	&	1.1946	&	0.0128	&	0.3430	&	0.0008	&	0.1238	&	0.0009	\\
57537.707450	&	2.07	&	1.26	&	1.5314	&	0.0138	&	0.3650	&	0.0009	&	0.1350	&	0.0010	\\
57538.662184	&	-2.16	&	1.28	&	1.6772	&	0.0159	&	0.3841	&	0.0008	&	0.1372	&	0.0010	\\
57540.697701	&	-6.62	&	1.00	&	1.5303	&	0.0131	&	0.3708	&	0.0007	&	0.1328	&	0.0008	\\
57549.682126	&	4.74	&	1.68	&	1.3937	&	0.0170	&	0.3498	&	0.0007	&	0.1220	&	0.0010	\\
57552.625769	&	-3.35	&	1.31	&	1.2904	&	0.0137	&	0.3494	&	0.0009	&	0.1277	&	0.0010	\\
57553.657209	&	-0.22	&	1.99	&	1.2436	&	0.0204	&	0.3455	&	0.0012	&	0.1438	&	0.0015	\\
57594.557626	&	-8.86	&	1.22	&	1.3356	&	0.0137	&	0.3453	&	0.0008	&	0.1273	&	0.0009	\\
57596.591265	&	-5.61	&	1.68	&	1.2850	&	0.0221	&	0.3466	&	0.0012	&	0.1268	&	0.0014	\\
57597.552753	&	-3.94	&	1.27	&	1.3464	&	0.0161	&	0.3459	&	0.0009	&	0.1219	&	0.0010	\\
57603.557020	&	2.35	&	1.28	&	1.3456	&	0.0132	&	0.3474	&	0.0008	&	0.1274	&	0.0009	\\
57604.544702	&	1.77	&	1.50	&	1.2943	&	0.0167	&	0.3497	&	0.0010	&	0.1246	&	0.0012	\\
57605.641977	&	-0.07	&	2.15	&	1.3678	&	0.0295	&	0.3528	&	0.0012	&	0.1246	&	0.0017	\\
57606.550501	&	0.41	&	1.42	&	1.3565	&	0.0183	&	0.3519	&	0.0010	&	0.1294	&	0.0012	\\
57607.570725	&	6.29	&	1.61	&	1.4251	&	0.0218	&	0.3557	&	0.0010	&	0.1252	&	0.0013	\\
57608.638735	&	0.77	&	1.01	&	1.4773	&	0.0180	&	0.3574	&	0.0009	&	0.1293	&	0.0011	\\
57609.585843	&	0.97	&	2.28	&	1.3597	&	0.0389	&	0.3549	&	0.0016	&	0.1320	&	0.0022	\\
57620.487606	&	-0.29	&	1.83	&	1.6431	&	0.0236	&	0.3570	&	0.0013	&	0.1356	&	0.0015	\\
57621.519280	&	-0.51	&	2.12	&	1.2062	&	0.0324	&	0.3524	&	0.0015	&	0.1353	&	0.0021	\\
57622.503754	&	-4.46	&	1.22	&	1.4621	&	0.0133	&	0.3542	&	0.0008	&	0.1299	&	0.0009	\\
57623.477128	&	-6.16	&	1.48	&	1.4835	&	0.0134	&	0.3562	&	0.0010	&	0.1328	&	0.0010	\\
57624.455987	&	-6.23	&	1.14	&	1.3930	&	0.0123	&	0.3510	&	0.0009	&	0.1299	&	0.0009	\\
57625.548799	&	-6.86	&	1.22	&	1.4067	&	0.0107	&	0.3501	&	0.0008	&	0.1309	&	0.0008	\\
57626.554737	&	-8.50	&	1.76	&	1.3159	&	0.0151	&	0.3482	&	0.0009	&	0.1280	&	0.0011	\\
57627.535108	&	-5.02	&	1.61	&	1.2625	&	0.0159	&	0.3474	&	0.0011	&	0.1282	&	0.0012	\\
57629.549641	&	-5.18	&	1.44	&	1.3152	&	0.0133	&	0.3471	&	0.0010	&	0.1248	&	0.0010	\\
57630.536443	&	-7.73	&	1.54	&	1.1276	&	0.0156	&	0.3402	&	0.0011	&	0.1245	&	0.0012	\\
57632.535169	&	-2.18	&	1.63	&	1.2713	&	0.0142	&	0.3436	&	0.0009	&	0.1269	&	0.0010	\\
57635.392487	&	-0.29	&	1.19	&	1.3157	&	0.0109	&	0.3477	&	0.0008	&	0.1270	&	0.0008	\\
57636.370440	&	-1.18	&	0.99	&	1.3496	&	0.0098	&	0.3481	&	0.0007	&	0.1302	&	0.0008	\\
57637.516115	&	1.77	&	1.12	&	1.3111	&	0.0108	&	0.3475	&	0.0009	&	0.1298	&	0.0009	\\
57638.545648	&	-2.87	&	1.45	&	1.3831	&	0.0135	&	0.3467	&	0.0009	&	0.1286	&	0.0010	\\
57640.538743	&	1.73	&	0.86	&	1.3627	&	0.0105	&	0.3503	&	0.0006	&	0.1242	&	0.0007	\\
57641.534605	&	3.05	&	1.03	&	1.3165	&	0.0122	&	0.3449	&	0.0008	&	0.1266	&	0.0009	\\
57643.537577	&	2.60	&	1.06	&	1.4228	&	0.0138	&	0.3515	&	0.0009	&	0.1305	&	0.0010	\\
57645.534503	&	-0.16	&	1.31	&	1.3778	&	0.0178	&	0.3520	&	0.0011	&	0.1279	&	0.0012	\\
57646.524307	&	-1.35	&	1.02	&	1.4791	&	0.0116	&	0.3541	&	0.0007	&	0.1280	&	0.0008	\\
57653.553135	&	1.14	&	1.43	&	1.6433	&	0.0218	&	0.3674	&	0.0011	&	0.1346	&	0.0013	\\
57654.555535	&	1.42	&	1.63	&	1.6412	&	0.0215	&	0.3653	&	0.0010	&	0.1348	&	0.0012	\\
57679.395908	&	3.04	&	1.16	&	1.4323	&	0.0128	&	0.3539	&	0.0007	&	0.1272	&	0.0009	\\
57680.409564	&	2.54	&	1.20	&	1.3705	&	0.0118	&	0.3533	&	0.0010	&	0.1297	&	0.0009	\\
57681.457267	&	-0.04	&	1.24	&	1.3597	&	0.0138	&	0.3547	&	0.0013	&	0.1346	&	0.0012	\\
57683.438943	&	-1.91	&	1.36	&	1.4271	&	0.0142	&	0.3532	&	0.0009	&	0.1293	&	0.0010	\\
57701.397060	&	-7.45	&	2.52	&	1.5908	&	0.0695	&	0.3524	&	0.0019	&	0.1311	&	0.0029	\\
57702.391756	&	-10.58	&	2.53	&	1.2207	&	0.0327	&	0.3521	&	0.0015	&	0.1285	&	0.0020	\\
57721.324810	&	-4.88	&	1.91	&	1.3428	&	0.0300	&	0.3470	&	0.0015	&	0.1299	&	0.0018	\\
57859.730190	&	-2.26	&	2.28	&	1.1723	&	0.0352	&	0.3477	&	0.0016	&	0.1290	&	0.0020	\\
57861.737737	&	1.98	&	1.35	&	1.3144	&	0.0160	&	0.3443	&	0.0009	&	0.1288	&	0.0011	\\
57865.717589	&	1.83	&	1.90	&	1.6984	&	0.0333	&	0.3519	&	0.0014	&	0.1330	&	0.0018	\\
57878.687660	&	5.46	&	1.24	&	1.4959	&	0.0184	&	0.3537	&	0.0009	&	0.1304	&	0.0012	\\
57880.690696	&	2.55	&	1.35	&	1.5076	&	0.0146	&	0.3580	&	0.0009	&	0.1329	&	0.0010	\\
57881.708813	&	1.11	&	1.21	&	1.4962	&	0.0229	&	0.3652	&	0.0012	&	0.1337	&	0.0014	\\
57890.711349	&	0.05	&	2.10	&	1.4537	&	0.0291	&	0.3562	&	0.0015	&	0.1368	&	0.0018	\\
57894.704442	&	1.82	&	1.19	&	1.4400	&	0.0122	&	0.3523	&	0.0008	&	0.1351	&	0.0009	\\
57896.703867	&	4.42	&	1.54	&	1.4497	&	0.0155	&	0.3520	&	0.0009	&	0.1324	&	0.0011	\\
57897.719003	&	2.20	&	1.31	&	1.4550	&	0.0157	&	0.3497	&	0.0009	&	0.1314	&	0.0011	\\
57898.725167	&	0.41	&	0.83	&	1.4378	&	0.0104	&	0.3495	&	0.0007	&	0.1322	&	0.0008	\\
57916.711750	&	5.64	&	1.14	&	1.3962	&	0.0139	&	0.3599	&	0.0011	&	0.1358	&	0.0011	\\
57928.702352	&	6.03	&	4.73	&	1.1069	&	0.0671	&	0.3393	&	0.0017	&	0.1242	&	0.0035	\\
57930.649599	&	10.56	&	2.14	&	1.2831	&	0.0230	&	0.3440	&	0.0009	&	0.1242	&	0.0014	\\
57932.666903	&	1.58	&	1.22	&	1.2986	&	0.0130	&	0.3400	&	0.0008	&	0.1281	&	0.0009	\\
57933.705004	&	0.59	&	1.26	&	1.3840	&	0.0144	&	0.3509	&	0.0009	&	0.1318	&	0.0010	\\
57934.663692	&	1.63	&	2.51	&	1.2205	&	0.0352	&	0.3404	&	0.0015	&	0.1291	&	0.0021	\\
57935.678133	&	0.74	&	2.21	&	1.2430	&	0.0230	&	0.3499	&	0.0012	&	0.1289	&	0.0015	\\
57944.717102	&	-4.60	&	1.36	&	1.2721	&	0.0128	&	0.3418	&	0.0007	&	0.1253	&	0.0009	\\
57952.634290	&	7.74	&	1.54	&	1.3957	&	0.0164	&	0.3541	&	0.0009	&	0.1327	&	0.0011	\\
57953.620628	&	8.78	&	1.11	&	1.4560	&	0.0109	&	0.3577	&	0.0007	&	0.1314	&	0.0008	\\
57969.481739	&	-1.55	&	1.42	&	1.2825	&	0.0184	&	0.3432	&	0.0011	&	0.1287	&	0.0013	\\
57972.527474	&	2.69	&	1.68	&	1.1616	&	0.0146	&	0.3417	&	0.0010	&	0.1271	&	0.0011	\\
57974.579433	&	-3.44	&	1.68	&	1.5073	&	0.0181	&	0.3527	&	0.0010	&	0.1360	&	0.0012	\\
57976.626596	&	-6.89	&	1.58	&	1.2432	&	0.0152	&	0.3437	&	0.0009	&	0.1259	&	0.0011	\\
57977.476387	&	-7.18	&	1.68	&	1.2844	&	0.0178	&	0.3485	&	0.0012	&	0.1312	&	0.0013	\\
57978.522875	&	-7.89	&	1.23	&	1.2845	&	0.0113	&	0.3427	&	0.0008	&	0.1275	&	0.0009	\\
57979.533008	&	-5.14	&	1.37	&	1.2524	&	0.0150	&	0.3413	&	0.0011	&	0.1294	&	0.0012	\\
57980.524437	&	-5.09	&	1.12	&	1.2436	&	0.0102	&	0.3434	&	0.0008	&	0.1260	&	0.0008	\\
57981.531616	&	-4.30	&	1.26	&	1.3445	&	0.0130	&	0.3425	&	0.0008	&	0.1273	&	0.0009	\\
57984.545703	&	-1.78	&	1.58	&	1.2999	&	0.0187	&	0.3441	&	0.0011	&	0.1240	&	0.0013	\\
57991.578297	&	16.01	&	3.53	&	1.6449	&	0.0664	&	0.3484	&	0.0014	&	0.1182	&	0.0026	\\
57993.584859	&	8.69	&	1.61	&	1.4253	&	0.0183	&	0.3506	&	0.0008	&	0.1259	&	0.0011	\\
57995.488618	&	3.24	&	1.78	&	1.4359	&	0.0197	&	0.3534	&	0.0010	&	0.1285	&	0.0013	\\
57997.489417	&	-1.74	&	1.44	&	1.3938	&	0.0202	&	0.3547	&	0.0011	&	0.1302	&	0.0013	\\
58000.496416	&	2.97	&	1.59	&	1.3261	&	0.0159	&	0.3527	&	0.0011	&	0.1291	&	0.0012	\\
58005.520203	&	-2.82	&	1.23	&	1.4496	&	0.0134	&	0.3561	&	0.0008	&	0.1278	&	0.0009	\\
58006.527222	&	-2.75	&	1.54	&	1.3786	&	0.0166	&	0.3547	&	0.0009	&	0.1222	&	0.0011	\\
58007.519612	&	0.12	&	1.57	&	1.3121	&	0.0161	&	0.3507	&	0.0009	&	0.1281	&	0.0011	\\
58008.508170	&	-1.55	&	1.33	&	1.3316	&	0.0121	&	0.3521	&	0.0008	&	0.1276	&	0.0009	\\
58009.518323	&	-1.39	&	3.10	&	1.3530	&	0.0494	&	0.3518	&	0.0018	&	0.1327	&	0.0025	\\
58010.527329	&	1.69	&	1.58	&	1.3843	&	0.0161	&	0.3481	&	0.0009	&	0.1264	&	0.0011	\\
58011.410126	&	-1.12	&	1.16	&	1.3190	&	0.0145	&	0.3504	&	0.0010	&	0.1330	&	0.0011	\\
58024.541278	&	0.95	&	1.54	&	1.3241	&	0.0185	&	0.3459	&	0.0009	&	0.1257	&	0.0011	\\
58025.371544	&	1.20	&	1.39	&	1.2873	&	0.0106	&	0.3445	&	0.0008	&	0.1268	&	0.0008	\\
58031.495137	&	0.25	&	1.69	&	1.4461	&	0.0153	&	0.3483	&	0.0009	&	0.1335	&	0.0011	\\
58037.441370	&	4.77	&	1.85	&	1.5957	&	0.0214	&	0.3530	&	0.0010	&	0.1317	&	0.0013	\\
58333.524276	&	7.28	&	1.61	&	1.3096	&	0.0121	&	0.3452	&	0.0009	&	0.1282	&	0.0010	\\
58334.519679	&	5.84	&	1.43	&	1.3670	&	0.0137	&	0.3413	&	0.0010	&	0.1314	&	0.0011	\\
58335.503783	&	8.15	&	4.77	&	2.0728	&	0.1667	&	0.3382	&	0.0038	&	0.1319	&	0.0075	\\
58443.358465	&	3.62	&	1.18	&	1.2693	&	0.0118	&	0.3384	&	0.0008	&	0.1258	&	0.0009	\\
58446.338876	&	-4.56	&	2.05	&	1.1359	&	0.0278	&	0.3387	&	0.0012	&	0.1284	&	0.0017	\\
58974.692775	&	2.03	&	1.56	&	1.2699	&	0.0143	&	0.3461	&	0.0009	&	0.1262	&	0.0011	\\
58976.693276	&	5.43	&	1.91	&	1.2408	&	0.0166	&	0.3468	&	0.0010	&	0.1265	&	0.0012	\\
58985.653439	&	1.51	&	1.99	&	1.3204	&	0.0154	&	0.3511	&	0.0009	&	0.1282	&	0.0011	\\
59007.691145	&	-0.18	&	1.32	&	1.1632	&	0.0090	&	0.3409	&	0.0008	&	0.1277	&	0.0008	\\
59050.687681	&	-4.26	&	1.52	&	1.2192	&	0.0120	&	0.3381	&	0.0009	&	0.1209	&	0.0009	\\
59051.675891	&	-2.99	&	1.51	&	1.2805	&	0.0135	&	0.3445	&	0.0007	&	0.1212	&	0.0009	\\
59068.490382	&	9.43	&	1.47	&	1.5259	&	0.0148	&	0.3536	&	0.0007	&	0.1263	&	0.0010	\\
59069.572791	&	-0.57	&	1.31	&	1.5124	&	0.0113	&	0.3565	&	0.0007	&	0.1306	&	0.0008	\\
59070.632248	&	-1.41	&	1.26	&	1.5447	&	0.0148	&	0.3515	&	0.0008	&	0.1308	&	0.0010	\\
59071.656784	&	-1.96	&	1.69	&	1.4222	&	0.0196	&	0.3489	&	0.0010	&	0.1301	&	0.0012	\\
59111.440202	&	0.82	&	1.62	&	1.3842	&	0.0147	&	0.3475	&	0.0012	&	0.1292	&	0.0012	\\
59125.467731	&	3.12	&	1.41	&	1.3467	&	0.0187	&	0.3399	&	0.0009	&	0.1223	&	0.0012	\\
59126.368868	&	-1.78	&	1.46	&	1.2907	&	0.0120	&	0.3401	&	0.0007	&	0.1220	&	0.0009	\\
59130.372594	&	-8.46	&	2.44	&	1.3283	&	0.0269	&	0.3395	&	0.0012	&	0.1219	&	0.0016	\\

\hline\noalign{\smallskip}
\end{longtable}
}%


\section{Online figures}

%
\begin{figure}[!htb]
\centering
\includegraphics[scale=0.65]{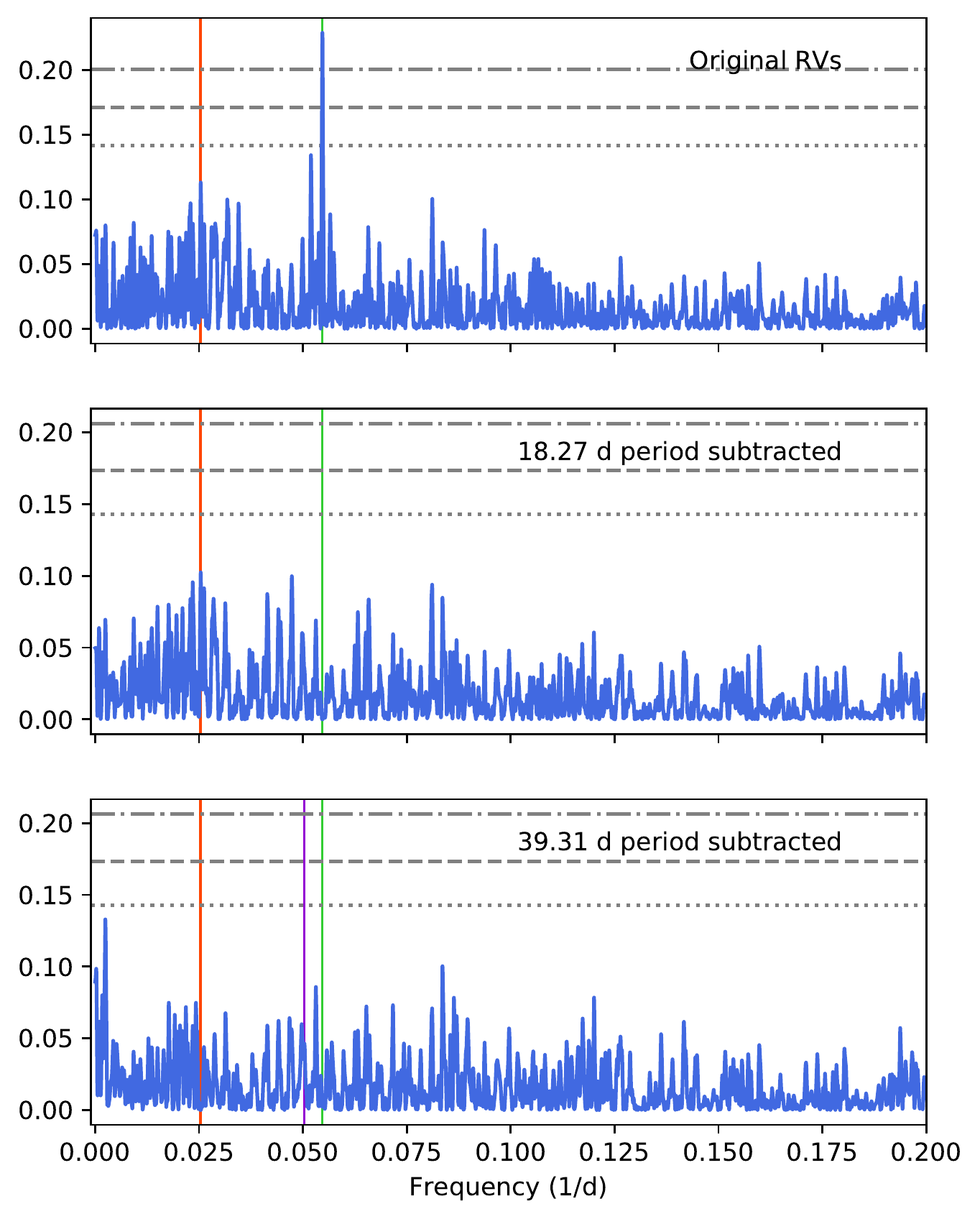}
\caption{ Top: GLS periodogram of the DRS RV measurements.
Middle: GLS periodogram after subtracting the 18.27 d period.
Bottom: GLS periodogram after subtracting the 18.27 and the 39.31 d signals. Values
corresponding to a FAP of 10\%, 1\%, and 0.1\% are shown with
horizontal grey lines. The vertical red line indicates the period at 39.31 d
while the vertical green line shows the 18.27 d period.
The first harmonic of the 39.31 d signal is shown in violet.
} 
\label{rv_series_periodogram_drs}
\end{figure}

\begin{figure*}[!htb]
\centering
\begin{minipage}{0.48\linewidth}
\includegraphics[scale=0.50]{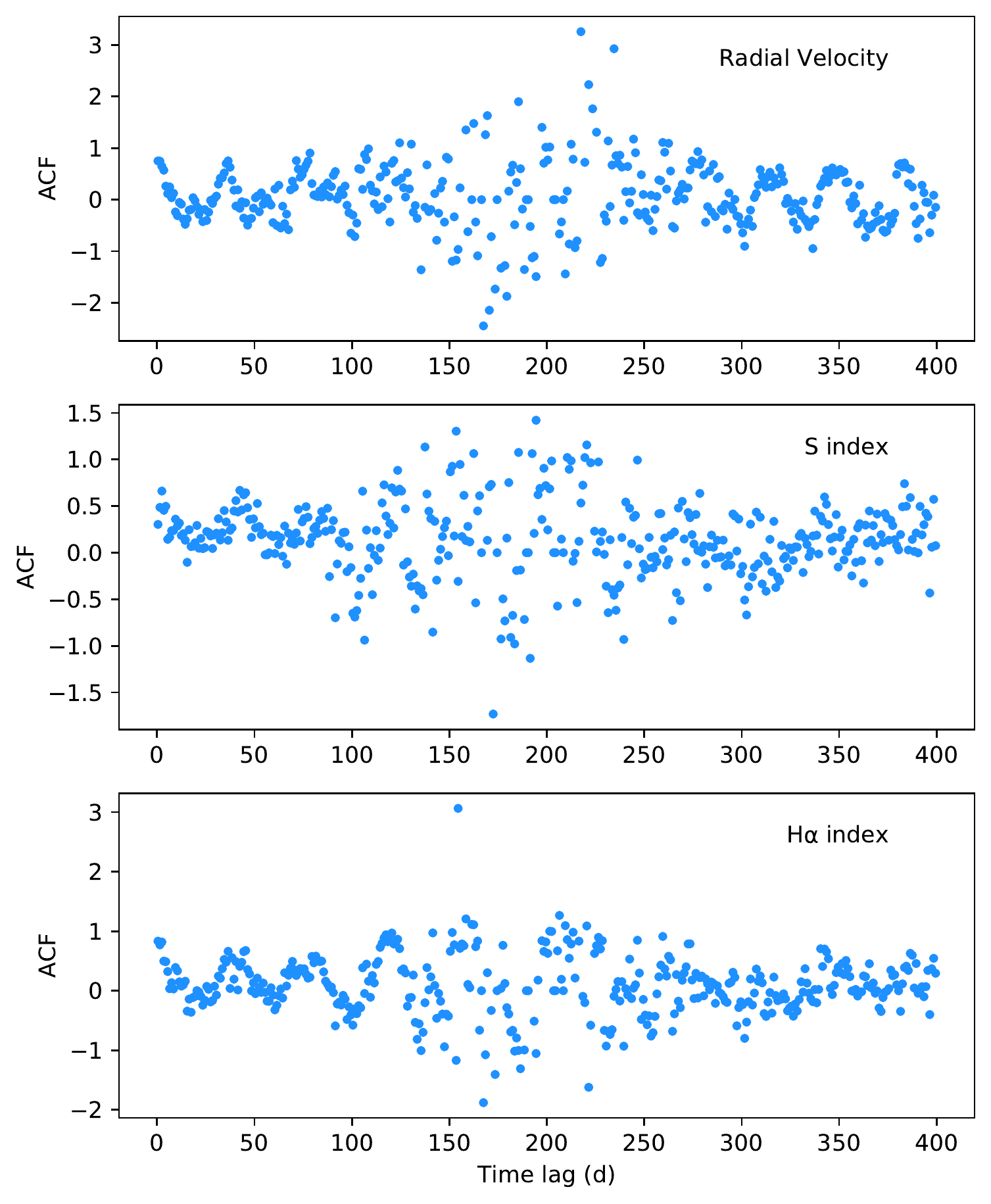}
\end{minipage}
\begin{minipage}{0.48\linewidth}
\includegraphics[scale=0.55]{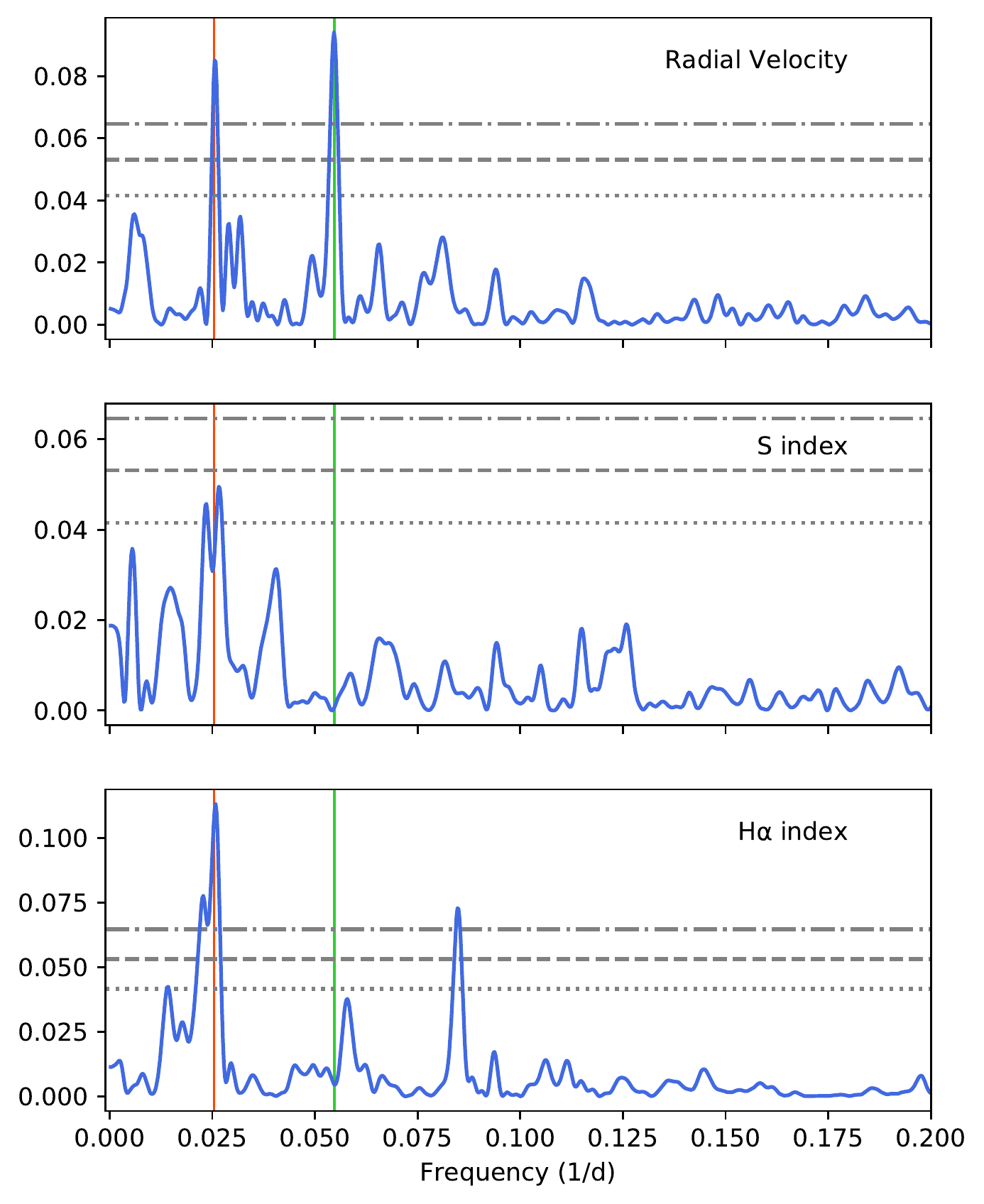}
\end{minipage}
\caption{
Left: autocorrelation function of RV, S-index, and H$\alpha$ time-series. Right: Corresponding GLS diagrams.}
\label{acf_functions}
\end{figure*}

\begin{figure*}
\includegraphics[width=2\columnwidth]{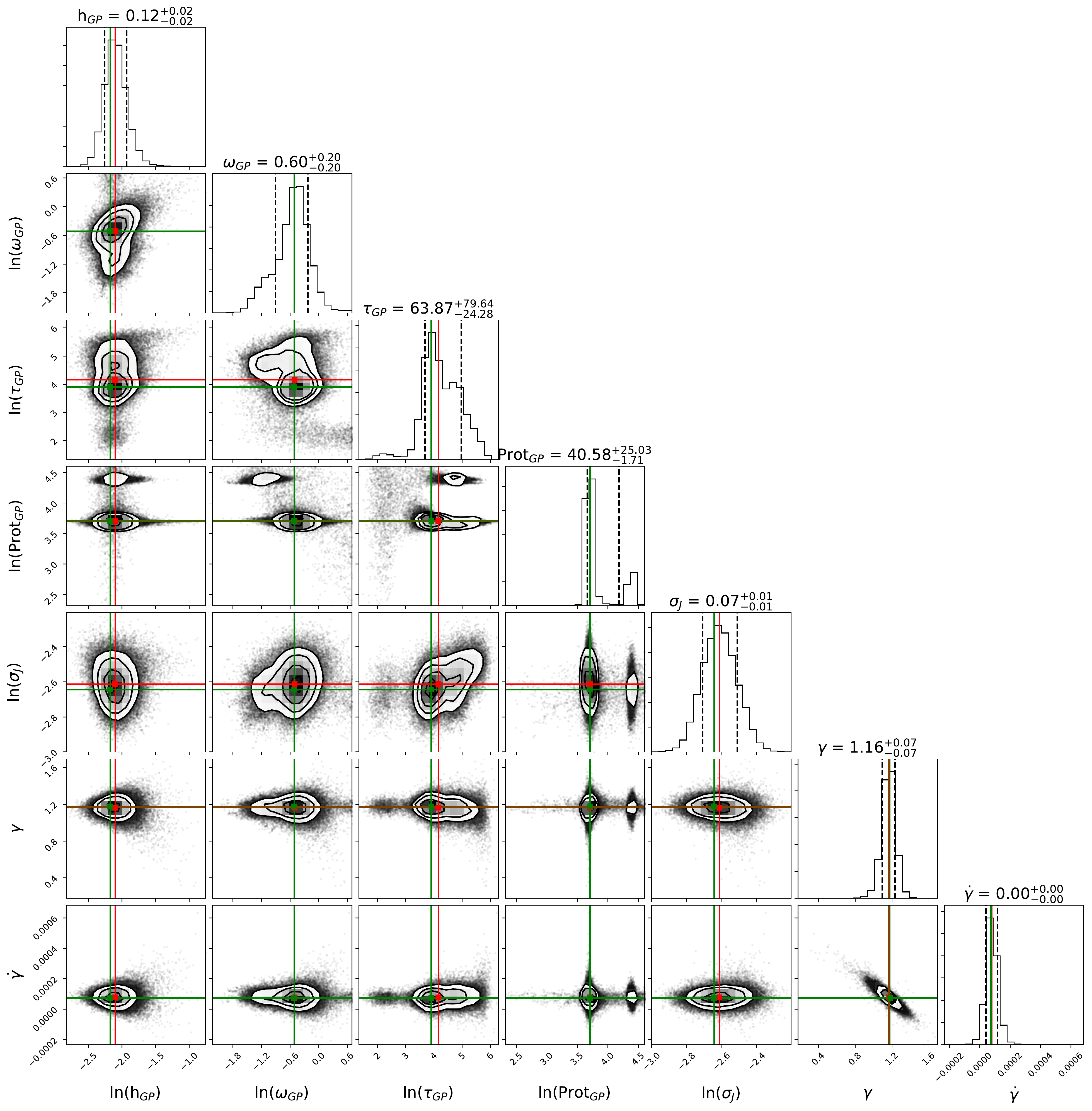}
\caption{Posterior distribution of the 'GP-only' model of the S-index time series in which median and maximum a-posterior probability (MAP) have been marked (respectively, red and green line).}
\label{star_only_activity}
\end{figure*}

\begin{figure*}
\includegraphics[width=2\columnwidth]{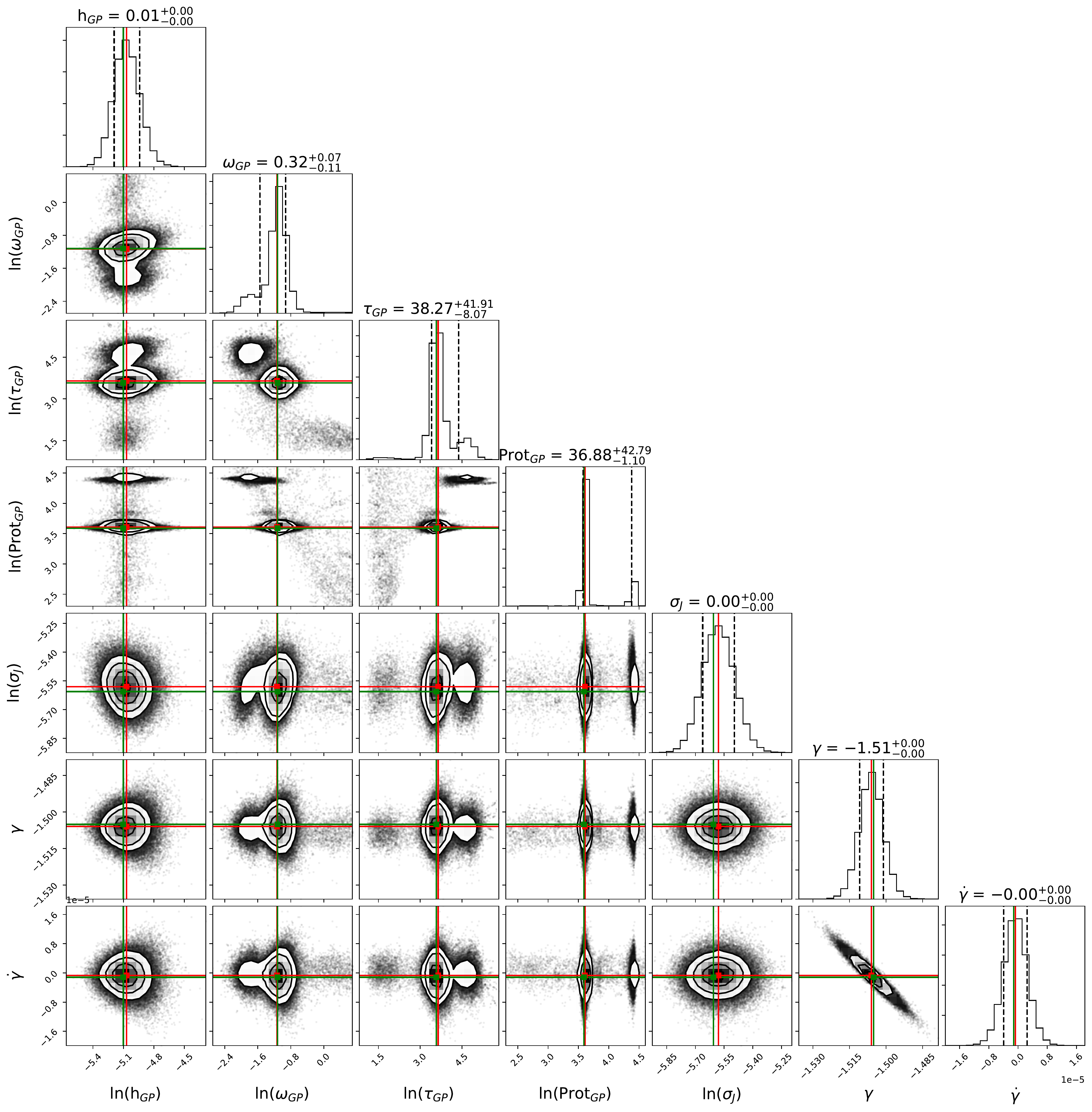}
\caption{Posterior distribution of the 'GP-only' model of the EXORAP V band photometry in which median and maximum a-posterior probability (MAP) have been marked (respectively, red and green line).}
\label{star_only_photometry}
\end{figure*}

\end{appendix}

\end{document}